\documentclass[a4paper]{amsart}
\usepackage{epsfig}
\usepackage{oldlfont}

\newtheorem*{ack}{Acknowledgment}
\theoremstyle{plain}
\newtheorem{thm}{Theorem}

\newcommand{\ev}[1]{\left\langle #1 \right\rangle}
\newcommand{\bra}[1]{\langle #1 \mid}
\newcommand{\ket}[1]{\mid #1 \rangle}
\newcommand{\braket}[2]{\langle #1 \mid #2 \rangle}

\newcommand{\pic}[5]{\raisebox{#3pt}
{\hspace{#4pt} \epsfig{file=#1.ps,height=#2pt,silent=} 
\hspace{#5pt}}}
\newcommand{\kd}[1]{\mathchoice{
\pic{#1}{24}{-8}{-1}{2}}{
\pic{#1}{11}{-3}{1}{1}}{
\pic{#1}{9}{-2}{-3}{1}}{
\pic{#1}{7}{-1}{-1}{0}}}

\newcommand{\lkd}[1]{\mathchoice{
\pic{#1}{40}{-18}{-1}{2}}{
\pic{#1}{25}{-11}{1}{1}}{
\pic{#1}{20}{-10}{-3}{1}}{
\pic{#1}{18}{-8}{-1}{0}}}

\begin{document}

\title{Operators for quantized directions}

\author{Seth A. Major}

\date{27 August 1999}
\address{Institut f\"ur Theoretische Physik 
\\ Der Universit\"at Wien \\ Boltzmanngasse 5\\
A-1090 Wien AUSTRIA}
\email{smajor1@swarthmore.edu (current)}

\begin{abstract}
Inspired by the spin geometry theorem, two operators are defined which 
measure angles in the quantum theory of geometry.  One operator 
assigns a discrete angle to every pair of surfaces passing through a 
single vertex of a spin network.  This operator, which is effectively 
the cosine of an angle, is defined via a scalar product density 
operator and the area operator.  The second operator assigns an angle 
to two ``bundles'' of edges incident to a single vertex.  While somewhat more 
complicated than the earlier geometric operators, there are a number 
of properties that are investigated including the full spectrum of 
several operators and, using results of the spin geometry theorem, 
conditions to ensure that semiclassical geometry states replicate 
classical angles.

\end{abstract}

\maketitle

\begin{flushright}
		UWThPh - 1999 - 28
\end{flushright}

\section{Introduction}

Spin networks were first used by Penrose as a combinatorial foundation for 
Euclidean three-space \cite{penrose}.  When first defined, spin 
networks were non-embedded, trivalent graphs with spins assigned to 
every edge.  Together with Moussouris, Penrose constructed a proof 
which demonstrated that the angles of three-dimensional space could be 
modeled by spin networks.  This proof rests on conditions on the form 
of semiclassical states.  They must be sufficiently correlated and the 
edges must have large spins.  Penrose called this result the ``spin 
geometry theorem.''

In 1994 spin networks were shown to be useful in the study of 
non-perturbative, canonical quantum gravity \cite{CLSpin}.  (See Ref.  
\cite{CRrev} for a recent review.)  Since then, spin networks have 
become a key element in the kinematics of the theory.  In fact, spin 
networks are the eigenvectors of operators which measure geometric 
quantities such as area and volume \cite{RSareavol} - \cite{QGII} and 
are a basis for the (kinematical) states of quantum gravity \cite{B}.  
This collection of work is often described as loop quantum gravity or, 
emphasizing the kinematic level, quantum geometry \cite{QGI}.  Given 
these two spin network developments - the spin geometry theorem and 
the introduction of spin networks to quantum gravity - one might 
expect that there is a well-defined ``angle operator'' for quantum 
geometry.  Such an operator exists and is defined in this paper.

In fact, I introduce several operators, two of which maybe be called 
``angle operators'' and both of which directly lead to ``quantized 
directions.''  For each of these operators I  
define the quantum operators, compute the spectra, and then check the 
naive classical limit and construct a regularization.  In so doing, 
the physical meaning and the interpretation of these operators becomes 
clear.  The first cosine operator is defined in two stages.  First, a 
scalar product density operator is introduced.  Second, the scalar 
product operator is normalized by the point-wise areas of the two 
surfaces.  The resulting operator - which is seen to give the cosine 
of the angle between the two surface normals - is a well-defined, 
self-adjoint operator on the space of kinematic states of quantum 
gravity.  The second cosine operator and the associated angle operator 
are constructed with similar techniques but are based on 
``orthogonal'' surfaces.

These steps are similar to the development of the ``cosine operator'' in 
Moussouris' dissertation \cite{JM}.  Since this work is unpublished, 
it is worth reviewing this construction in some detail.  This is done 
in Section \ref{spingeom}.  In Section \ref{qg} there is a brief 
review of quantum geometry as it has developed in the background 
independent quantization of Hamiltonian gravity.  In Section \ref{aop} 
a scalar product density operator is introduced and the spectrum 
computed.  Then the cosine operator is defined.  This operator is 
shown to have the expected naive classical limit in Section 
\ref{classlim}.  There are two regularizations sketched in Section 
\ref{reg}.  In Section \ref{jetop} the second angle operator is 
introduced.  Some variations on the operator and the semiclassical 
limits are discussed in the final section of the paper.  Both of the 
operators share some striking features including a completely discrete 
spectra and independence of both the Planck length and the Immirizi 
parameter (\cite{I} - \cite{GOPI}).

\section{The Spin Geometry Theorem}
\label{spingeom}

Difficulties inherent in the continuum formulation of physics -- from 
ultra-violet divergences in quantum field theory to the evolution of 
regular data into singularities in general relativity -- led Penrose 
to explore a fundamentally discrete structure for spacetime.  His 
insight was that one could define the notion of direction with  
combinatorics of spin networks and recover the continuum of angles to 
arbitrary accuracy.  He accomplished this by using the discrete spectrum of 
angular momentum operators.

Relative orientations arise out of a spin network structure through 
scalar products of angular momentum operators.  The construction 
offers a way to determine angles in three dimensional space without 
any reference to background manifold structure.\footnote{The angle 
operator defined in quantum geometry does depend on the manifold 
structure through a dependence on the tangent space at vertices.  See 
Section \ref{aop}.} Realistic models of angles must be arbitrarily 
fine and are constructed with complex networks.

\begin{figure}
\begin{center}
\begin{tabular*}
{\textwidth}{c@{\extracolsep{\fill}}
c@{\extracolsep{\fill}}
c@{\extracolsep{\fill}}}
\pic{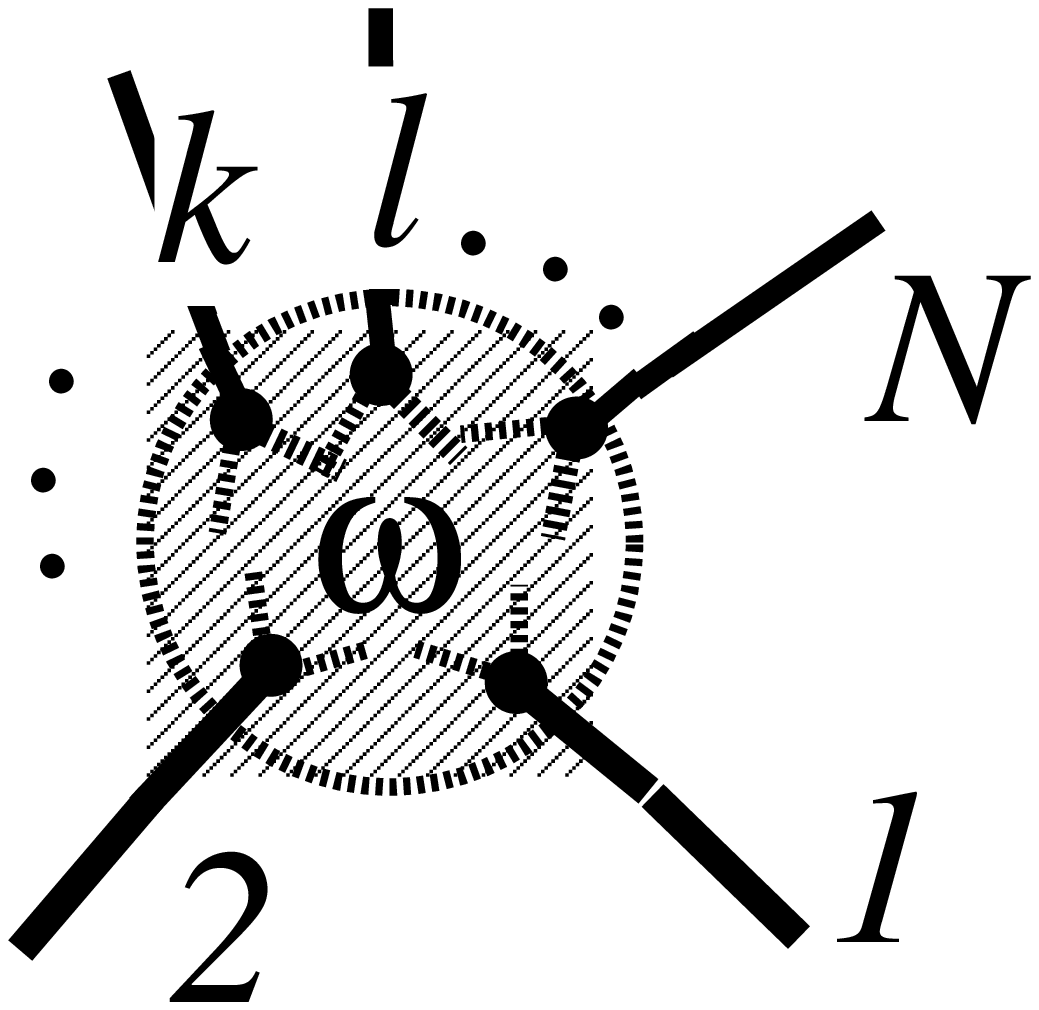}{55}{-5}{0}{0} & \pic{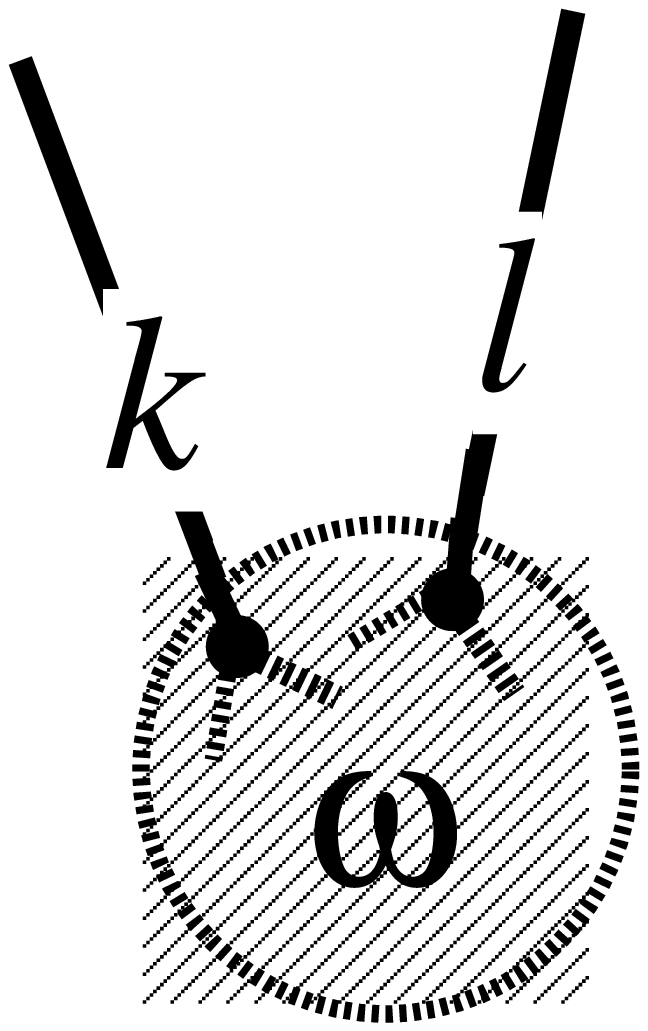}{55}{-5}{0}{0}  &
\pic{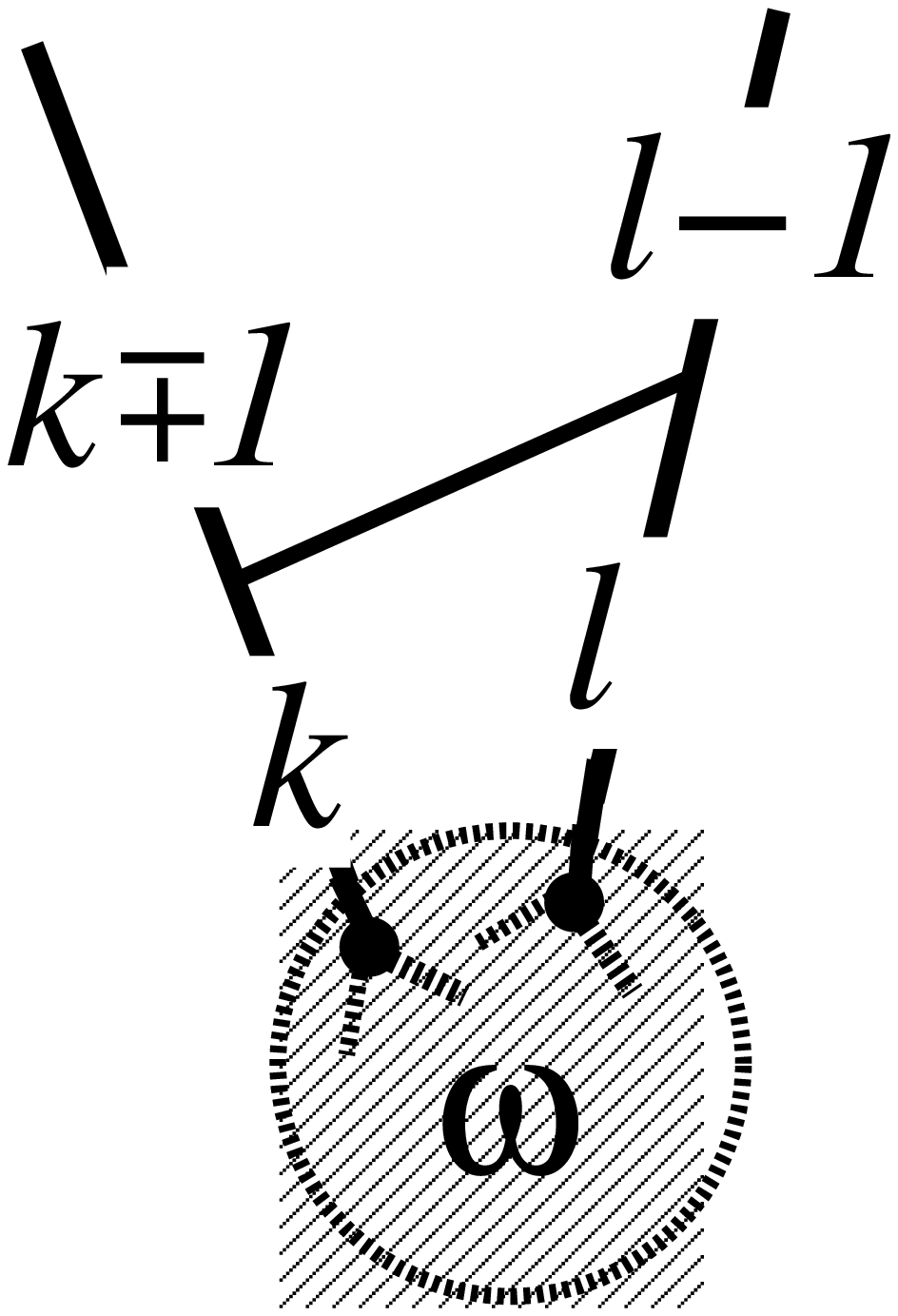}{70}{-5}{0}{0} \\
a & b & c\\
\end{tabular*}
\end{center}
\caption{ (a.)  A spin network state with $N$ external lines based on 
the invariant $\omega$; a spin network with only these $N$ open lines.  
The lines are labeled $1,2, \dots, N$.  Two of the spins $k$ and $l$ 
are identified.  (b.)  A particular example with two lines of $k$ and 
$l$ spin.  (c.)  The exchange of a spin-$1/2$ ``particle.''  This 
``experiment'' helps determine the angle between the two lines.}
\label{snkl}
\end{figure}

To see how this comes about, consider a spin state $\omega$ with $N$ 
correlated external lines as shown in Fig. (\ref{snkl}a).
These lines are built of $N$ ($N \ge 3$) spins $s_i$, $i=1,2, \dots 
N$.  The relative angles between the different lines are described by 
angular momentum operators $\hat{J}_{(k)}$ which act on the $k$th line 
of the graph.  (The indices in parentheses distinguish them from the 
indices of the spatial manifold.)  The scalar product of two such spin 
operators is given by $\hat{T}^{(kl)}$
\begin{equation}
\label{SPop}
\hat{T}^{(kl)} := \hat{J}_{(k)} \cdot \hat{J}_{(l)} \equiv
\sum_{i=1}^3 \left( 1 \otimes \dots \otimes \hat{J}^{i}_{(k)} \otimes 
\dots
\otimes \hat{J}^{i}_{(l)} \otimes \dots \otimes 1 \right).
\end{equation}
This operator acts non-trivially only on the two lines with spins 
$j_{k}$ and $j_{l}$.  For instance, if $\hat{J} = \hat{J}_{(1)} + 
\hat{J}_{(2)}$, then the operator $\hat{T}^{(12)}$ may be written as
\begin{equation}
	\label{T12op}
\hat{T}^{(12)} = \tfrac{1}{2} \left[ \hat{J}^2 - \hat{J}_{(1)}^2 - 
\hat{J}_{(2)}^2 \right]
\end{equation}
and has eigenvalues $\tfrac{1}{2} \left[ j(j+1) - j_1(j_1+1) - j_2(j_2+1)
\right]$.  (Throughout this paper I denote scalar products 
with $T^{(\dots)}$.)

The spin geometry theorem states that for a sufficiently classical 
state $\omega$ the expectation values $\bra{\omega} \hat{T}^{(kl)} 
\ket{\omega} \equiv \ev{\hat{T}^{(kl)}}_{\omega}$ model the scalar 
products of vectors in Riemannian 3-dimensional space.  For 
states $\omega$ which are the direct product of unique polarization 
vectors the expectation value $\ev{\hat{T}^{(kl)}}_{\omega}$ is 
precisely the inner product of those polarization vectors \cite{JM}.

In more detail, the interpretation of $\hat{T}^{(kl)}$ as scalar 
product of vectors requires a certain richness in the state $\omega$.  
Just as one must specify a set of conditions to find the Newtonian 
limit of general relativity, one must specify a set of classical limit 
conditions for these operators.  These are the ``classical constraints.''  
In the spin geometry theorem one has a choice of constraints based on 
the following scalar product lemma \cite{JM}:  If $T^{(kl)}$ is a 
real, symmetric $N \times N$ matrix then these three conditions are 
equivalent:
\begin{enumerate}
\item There exist 3-dimensional vectors $\{ 
\vec{v}\,{}^k \}$, $i=1,2, \dots N$,
such that $T^{(kl)}$ is the scalar product, i.e. 
$T^{(kl)} = \vec{v}\, {}^k \cdot \vec{v}\,{}^l$.
\item $T^{(kl)}$ is positive, semi-definite of rank $\leq 3$.
\item $ x_k \, T^{(kl)} \, x_l \ge 0$ for real $x_k$ and
the determinants of all symmetric $4 \times 4$ submatrices of 
$T^{(kl)}$ vanish.
\end{enumerate}
The proof is an application of linear algebra \cite{JM}.

In the quantum theory and for spins of finite magnitude, the classical 
constraints are only satisfied approximately.  Hence, one says that a 
matrix $\ev{\hat{T}^{(kl)}} = T^{(kl)}$ satisfies the 
``$\epsilon$-constraints'' if, for some $\epsilon > 0$,
\begin{enumerate}
\item  $T^{(kk)} > 0 $ and $x_k T^{kl}x_l 
\geq 0$ for real $x_k$ and
\item The matrix of cosines 
\[
C^{(kl)} =  \frac{ T^{(kl)} }{2 \left[ T^{(kk)} T^{(ll)}
\right]^{1/2} }
\]
satisfies
\[
\left| \det C^{(kl)} \right|^{1/2} < \epsilon
\]
when summing over $k,l$ in a 4-tuple of indices $K$.  This condition
requires that the four volume, defined by $K$, is less than $\epsilon$.
\end{enumerate}
By the properties of symmetric matrices given above, spin operators 
$\hat{T}^{(kl)}$ satisfying these classical constraints approximate 
condition 3 above for real, symmetric matrices.  This is how 
classical angles are approximated.

Quantum states can support this classical angle interpretation only when 
they are sufficiently correlated so that the geometric relations 
between spins give well-defined $\hat{T}^{(kl)}$'s.  This condition is 
specified using a bound on the uncertainty.  Defining the 
root-mean-square uncertainty in a state $\omega$ as
\[
\sigma_\omega \hat{T}^{(kl)} := \left[ \ev{ \left( \hat{T}^{(kl)} -
\ev{\hat{T}^{(kl)}}_\omega \right)^2 }_\omega \right]^{1/2},
\]
$\hat{T}^{(kl)}$ is ``$\delta$-classical'' in the state $\omega$ when
\[
\frac{ \sigma_\omega \hat{T}^{(kl)} } { \| \hat{T}^{(kl)} \| } < \delta
\]
where $\| \hat{T}^{(kl)} \| := {\rm sup} \{ |
\ev{\hat{T}^{(kl)}}_\omega| :
\| \omega \| = 1 \}$. (For finite spins $\hat{T}^{(kl)}$ is a bounded 
operator, as can be seen from Eq. (\ref{T12op}).) 
Since $\| \hat{T}^{(kl)}\|$ obtains the maximum value $j_k
j_l$, the uncertainty $\sigma_\omega
\hat{T}^{(kl)} / j_k j_l
\leq \delta$.  Thus, when the spins are large, the spin product 
operator $\hat{T}^{(kl)}$ models angles in 3-dimensional space.

The theorem is stated as
\begin{thm}{ \bf Spin Geometry Theorem:}
For all $\epsilon > 0$, there exists a $\delta >0$ such that the 
values $\ev{\hat{T}^{(kl)}}_\omega$ satisfy the $\epsilon$-constraints 
for Riemannian 3-space, provided the observable 
$\hat{T}^{(kl)}$ is $\delta$-classical in the state 
$\omega$.
\end{thm}
The proof may be found in Ref.  \cite{JM} and only rests on the 
assumption that $\omega$ contains enough information to be 
$\delta$-classical (and the parameter $\delta$ is independent of the 
state $\omega$).  In short, three dimensional angles are obtained if 
the state has two properties: Its spins must be large so the scalar 
products are fine enough to obtain the classical limit.  And the state 
must be sufficiently correlated so there is enough information to 
separate random correlation from relative orientation.

Penrose proves a similar result using diagrammatic techniques which I 
sketch here.  (See Refs.  \cite{penrose} and \cite{JM} for more 
detail.)  One can consider a network with two free ends as in Fig.  
(\ref{snkl}b).  Penrose builds a new network by splitting off one line 
from the $s_{l}$ line and connecting it to the $s_{k}$ line.  The two 
outcomes ($k\pm1$) are shown in Fig.(\ref{snkl}c).  The cosine of the 
angle between the two edges $k$ and $l$ is defined to be the relative 
probability of the two outcomes.  However, this is not sufficient.  
This angle is not the scalar product operator but also includes an 
``ignorance'' factor.  For instance, if the state $\omega$ was a set 
of uncorrelated lines then, in the limit of large spins, the two 
relative probabilities would become equal.  If any angle was assigned 
in this case, it would have to be a right angle.  Penrose fixes this 
ambiguity by making two successive measurements.  If the resulting 
angles are approximately the same, then Penrose says the angle is well 
defined.  Thus the cosine may be determined 
with
$$
\lkd{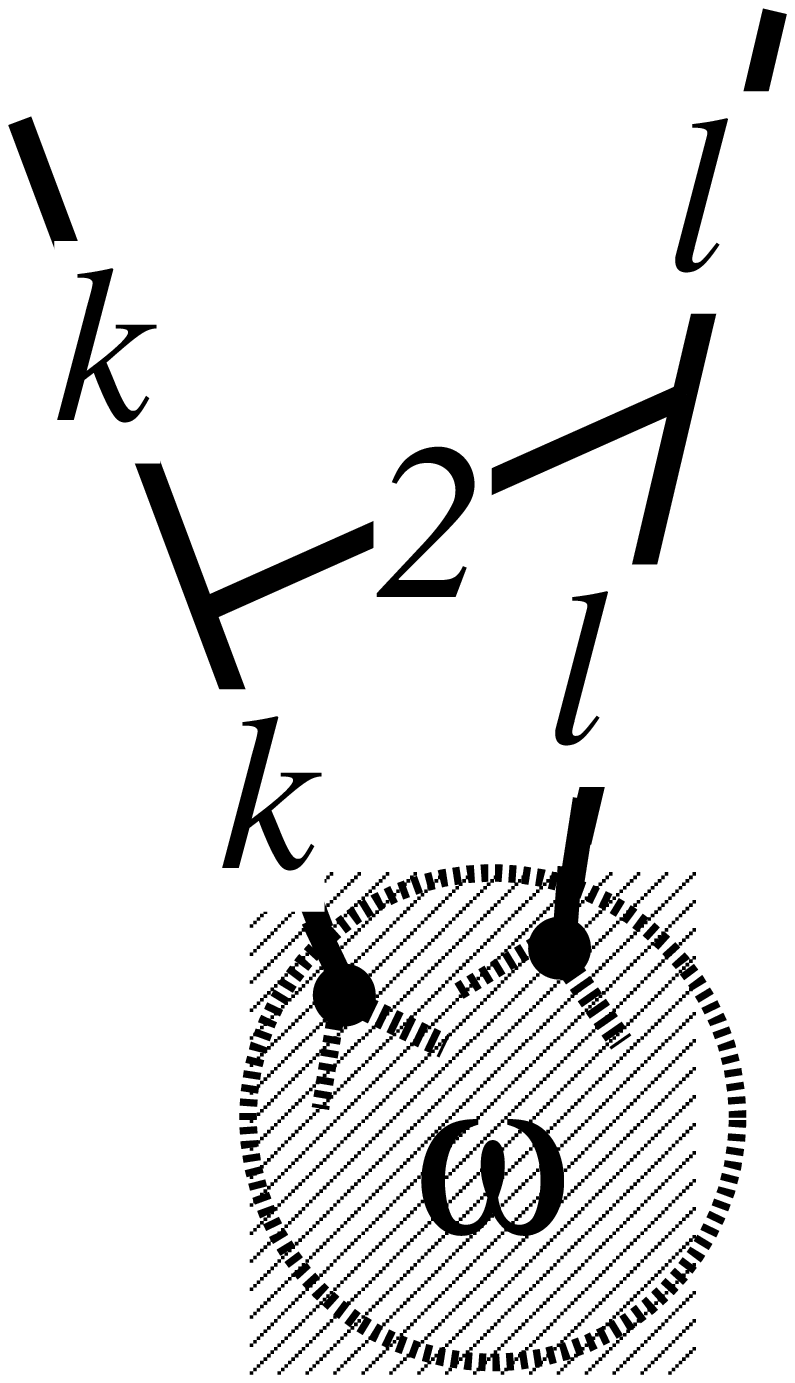} \approx k l \cos \theta \lkd{snkl1}.
$$
The cosine may be calculated using recoupling theory. The relation 
between this approach and the scalar product can be seen rather directly.
In brief, since
$J^{i} = \tfrac{\hbar}{2} \sigma^{i}$ where $\sigma^{i}$ are the Pauli 
matrices and since \cite{roberto}
\begin{equation}
	\label{pauli}
\frac{1}{2} \sum_{i=1}^{3} \sigma^{i \, B}_{A} 
\sigma^{i \, D}_{C}  \equiv \kd{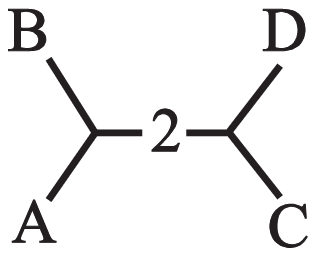}
\end{equation}
the angle is defined with a diagram identical to the 
one used in Eq. (\ref{SPop}). Indeed, the angle is an
eigenvector of the scalar product operator.    

The spin geometry theorem shows that it is possible to build a 
classical-looking angle on a fundamentally combinatorial space.  It is 
the limit which allows a fundamentally discrete spacetime to have 
classical properties.  In this same manner, the spin geometry theorem 
offers lessons for the current formulation of quantum gravity.  While 
the kinematic formulation is well understood and on rigorous footing, 
there is little notion of how to recover our familiar Minkowski 
spacetime.  The subtleties encountered in the Spin Geometry Theorem 
surely have a reflection in the classical limit of non-perturbative 
quantum gravity.

With the spin geometry theorem as motivation, this paper introduces 
operators which lead to quantized directions.  The idea is to, 
directly as possible, define angle operators for quantum geometry 
as they are defined in the spin geometry theorem.  As the setting for 
these operators is quantum geometry, this is reviewed first.

\section{Quantum Geometry: A brief review}
\label{qg}

In addition to a quick review of quantum geometry this section serves 
two purposes.  It sets the framework of the operator definitions and 
serves to fix signs, factors, and units.

The quantum geometry framework is suitable for a class of gauge 
theories which are quantized with canonical, background metric-free 
methods.  The notable example of such a theory is, of course, 
canonical quantum gravity.  The kinematics of this theory is placed on 
an oriented, analytic three manifold $\Sigma$ which is compact (or the 
fields to be mentioned shortly satisfy appropriate asymptotic 
conditions).  The classical configuration space $\cal{A}$ consists of 
all $su(2)$-valued smooth connections $A_{a}^{i}(x)$ on $\Sigma$.  
(The index $a$ is a one-form index while the $i$ is an internal 
Lie-algebra index.)  The phase space is the cotangent bundle over 
$\cal A$ with momenta represented by vector densities $E^{ai}(x)$ of 
weight one -- ``triads'' for short.  These two variables are conjugate
\begin{equation}
	\begin{split}
		\label{PB}
		\{A_{a}^{i}(x), A_{b}^{j}(y) \} = 0; \; & \; \; 
		\{E^{ai}(x), E^{bj}(y) \} = 0; \\
		\{A_{a}^{i}(x), E^{bj}(y) \} &= 8 \pi G \delta_{a}^{b} \delta^{ij} 
		\delta^{3}(x,y).
	\end{split}
\end{equation}
When quantizing, on account of the gauge and diffeomorphism invariances of
the theory, it is useful to construct configuration variables from
Wilson loops or holonomies of paths \cite{JS, RSloop, AI}.  To model 
geometry it is necessary to include vertices and so the state space
is most appropriately constructed out of graphs.  I denote
a graph embedded in $\Sigma$ by ${\mathsf G}$.  It contains a set of 
$N$ edges ${\bf e}$ and a set of vertices ${\bf v}$.  
Every connection
$A$ in ${ \cal A}$ associates a group element to an edge $e$ of 
$\bf e$ via the holonomy,
\[
U_e(A) := {\cal P} \exp  \int_e dt \dot{e}^a A_a.
\]
Here, $A_{a} := A_{a}^{i} \tau^{i}$ with $\tau^{i}$ proportional to 
the Pauli matrices via $\tau^{i} = - \tfrac{i}{2} \sigma^{i}$.
Every complex-valued smooth function of $N$ copies of the group
gives a function on ${\cal A}$
\[
C_{\mathsf G}(A) := c(U_{e_1}(A), \dots, U_{e_N}(A)).
\]
While these configuration variables only capture information about 
finite dimensional spaces of the infinite dimensional space ${\cal A}$,
when all graphs are included the space is large enough to separate points
in ${ \cal A}$.  As in linear field theories, these functions are
called cylindrical functions.  Associated to a particular graph, these
are denoted ${\rm Cyl}_{\mathsf G}$.  The union of these spaces over
all graphs, ${\rm Cyl}$, is taken to be the configuration space.

In the quantum theory the configuration space is necessarily enlarged 
to the space of generalized connections $\overline{\cal A}$.  One of 
these generalized connections $\bar{A}$ assigns to every edge $e$ in 
$\mathsf G$ a group element $\overline{A}(e)$ in $SU(2)$ 
\cite{ALproject2}.  It turns out that there is a measure on this space 
induced by the Haar measure on the group.  The kinematic Hilbert space 
of states, ${\cal H}$, is given by the square integrable functions on 
this space \cite{ALproject2} - \cite{MM}.  Elements of the Hilbert 
space
\[
\Psi_{G,c}(\bar{A}) := \prod_{e=1}^N c(U_e)
\]
enjoy the scalar product defined using the Haar measure
\[
\braket{\Psi_{G,c}}{\Psi_{G,c'}} := \int_{[SU(2)]^{n}} d^{n}g \, 
\overline{c(U_{e_{1}},..., 
U_{e_{N}})} c'(U_{e_{1}}, ... , U_{e_{N}}).
\]
(This is defined for two functions based on the same graph.  To see that
this places no restriction on the scalar product, note that any cylindrical
function of a graph can be expressed on a larger graph by assigning trivial
holonomies to edges not in the smaller graph.)

There is also a Hilbert space of states of square integrable functions 
of the gauge invariant configuration space $\overline{\cal A/G}$ 
(defined through a projective limit of $\cal A/G$ \cite{ALproject1} 
\cite{ALproject2}).  Most of this work is in this Hilbert space 
denoted by $\cal H$.  There is a basis on this set of states, the spin 
network basis\cite{B}.  In this context a spin network ${\cal N}$ 
consists of the triple $(\mathsf{G}; {\bf i, n})$ of an oriented 
graph, labels on the vertices or ``intertwiners,'' and integer edge 
labels indexing the representation carried by the edge.  The 
corresponding spin net state $\ket{s}$ in $\cal H$ is defined in the 
connection representation as
\[
\braket{A}{s} \equiv 
\braket{A}{ {\mathsf G} \, {\bf  i \,  n} } = \prod_{v \in {\bf v}( 
{\mathsf G} ) }
i_v  \circ \otimes_{e \in e({\mathsf G})} U_{(n_{e})}[A].
\]
The intertwiners are invariant tensors on the group so these states 
are gauge invariant.

The action of the triads on the configuration space may be computed 
from the Poisson brackets.  However, as the triads are dual to pseudo 
two-forms, they most comfortably live on two surfaces, generally 
denoted by $S$.  (There are subtleties with working with surfaces with 
boundary \cite{M}, \cite{QGI}.)  The variables are
\begin{equation}
\label{Edef}
E_S^i = \int_S d^2 \sigma \, n_a(\sigma) E^{ai} (x(\sigma))
\end{equation}
in which $\sigma$ are coordinates on the surface and 
$n_a = \epsilon_{abc} \tfrac{dx^b}{d\sigma_1}\tfrac{dx^c}{d\sigma_2}$
is the normal. Using the Poisson brackets of Eq. (\ref{PB}) 
and general properties of holonomies, one may show that for
a cylindrical function $C_{\mathsf G}$ \cite{QGIII} \cite{roberto}
\begin{equation}
	\label{EPB}
	\begin{split}
		\{ C_{\mathsf G}, E^i_S \} 
		&= 8 \pi G \int_{S} d^{2} \sigma \int_{e} dt n_{a}(\sigma)
		\dot{e}^{a}(t) \delta^{3}(x(t), y(\sigma))
		\left[ U_{e}(0,t) \tau^{i} U_{e}(t,1)\right]_{m}^{n}
		\left[ C_{({\mathsf G}-e)}\right]_{n}^{m}
		\\
		&= 4 \pi G \sum_{I \in \{ {\mathsf G} \cap S \} } 
		\chi^{S}_{I} \left[ C_{ ( {\mathsf G} - e) } \right]_{n}^{m} \,
		\begin{cases}
			[U_{e} \tau^{i} ]_{m}^{n} & \text{ if the edge is incoming, } 
			\\[ 9 pt]
			[ \tau^{i} U_{e} ]_{m}^{n} & \text{ if the edge is outgoing.}
		\end{cases}
	\end{split}
\end{equation}

I have introduced a bit of new notation: The function $C_{({\mathsf 
G}-e)}$ is based on the original graph without the edge $e$.  The 
indices $m,n$ are matrix indices for the representation carried by the 
edge $e$.  The sum in the second line is over all intersections, $I$, 
between the graph and the surface.  (If the surface cuts through an 
edge, a bivalent vertex is added to the graph ${\mathsf G}$.)  The 
sign factor $\chi^{S}_{I}$ is $+1$ for edges $I$ with orientations 
aligned with the surface normal and $-1$ for edges with orientations 
oppositely aligned with the surface normal.

There are also two remarks to make.  First, the result is 
non-vanishing only when there is at least one intersection between the 
graph ${\mathsf G}$ and the surface $S$.  Second, the overall factor 
of $\tfrac{1}{2}$ can be seen to arise from a ``thickened surface'' 
regularization \cite{QGIII}.  On the boundary of $S$ the issue is more 
delicate and is conjectured to have a jet dependence \cite{M}.

It is convenient to express the triad's action in terms of left (or right)
invariant vector fields on the $I$th copy of the group, $X_{I}^{i}$.  
Thus,
\begin{equation}
	\{C_{\mathsf G}, E_{S}^{i} \} = 4 \pi G \chi^{S}_{I}
	X_{I}^{i} \cdot C_{\mathsf G}.
\end{equation}
I have used the summation convention for the label $I$; all edges in 
the intersection of the graph and the surface are included.  The 
handedness of the vector field is given by the orientation of the 
edge.  With this preparation, we may directly define the quantum 
operator\footnote{As Immirizi has emphasized, in the canonical 
transformation used to define the connection there is a family of 
choices generated by one non-zero, real parameter $\gamma$, 
${}^{\gamma}A_{a}^{i} = \Gamma_{a}^{i} - \gamma K_{a}^{i}, 
{}^{\gamma}E^{ai} = (1/ \gamma)E^{ai}$ \cite{I}.  Throughout this work 
$\gamma =1$.}
\begin{equation}
	\label{Eop}
	\hat{E}_{S}^{i} \cdot \Psi_{\mathsf G} (A) := i  l^{2}
	\chi^{S}_{I} X_{I}^{i} \cdot \Psi_{\mathsf G}(A)
\end{equation}
in which the scale $l^{2} := 4 \pi G \hbar$ is introduced ($c=1$).  
This operator is essentially self-adjoint in ${\cal H}$ \cite{QGI}, 
\cite{ALProj}.  It is also useful to introduce the angular momentum 
operators associated to an edge, $\hat{J}_{(e)}^{i} \equiv i \hbar 
X_{(e)}^{i}$, which satisfy the usual algebra
\[
\left[ \hat{J}_{(e)}^{i}, \hat{J}_{(e')}^{j} \right] = i 
\epsilon^{ijk} \delta_{(e \, e')} \hat{J}_{(e)}^{k}. 
\]
The $\delta$-function restricts the relation to one edge; $\hat{J}$ on
distinct edges commute.  The
diagrammatic form of this operator is the ``one-handed''
\begin{equation}
	\label{Ediagram}
\hat{E}_{S}^{i} = i l^{2} \chi^{S}_{I} 
\pic{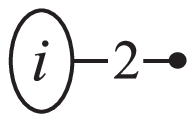}{14}{-6}{-1}{2}
\end{equation}
in which the index $i$ is the index of the angular momentum operator 
$\hat{J}^{i}$.  The grasping is chosen such that, in the plane of the 
digram, when the orientation on the edge points up, the 2-line is on 
the left (and on the right for downward orientations) \cite{RC}.  It 
is critical to note that the diagrammatic representation of a grasping 
involves the choice of a sign.  This is why the sign factor 
$\chi^{S}_{I}$ remains in the expression for the ``unclasped hand'' in 
Eq.  (\ref{Ediagram}).

The definition of the first cosine operator uses an operator of quantum 
geometry, the area.  I will review the construction here. For simplicity
let a surface $S$ be specified by $z=0$ in an adapted coordinate system.
Expressed in terms of the triad $E^{ai}$, the area of the surface 
only depends on the $z$-vector component via \cite{RSareavol}, 
\cite{FLR}, \cite{QGI}
\begin{equation}
	\label{areae}
	A_{S} = \int_{S} d^{2}x \sqrt{E_{z}^{i}  E_{z}^{i}}.
\end{equation}
The quantum operator is defined using the operators of Eq.  
(\ref{Eop}) and by partitioning the surface $S$ so that only one edge 
or vertex threads through each element of the partition.  Thus the 
integral of Eq.  (\ref{areae}) becomes a sum over operators only 
acting at intersections of the surface with the spin network.  With 
this ground work one may compute the spectrum.

The spectrum is most easily computed by first working with the square 
of the area operator.  Calling the square of the integrand of Eq.  
(\ref{areae}) $\hat{O}$, the two-handed operator at one intersection 
is
\begin{equation}
	\hat{O} \ket{s} =  (4 \pi G)^{2} 
	\chi^{S}_{I} \, \chi^{S}_J \,  \hat{J}_{I} \cdot
	\hat{J}_{J} \ket{s}.
\end{equation}
Here, $\hat{J}_I$
denotes the vector operator $\hat{J}$
acting on the edge $e_{I}$.
This operator is {\em almost} the familiar 
$\hat{J}^{2}$ but for the sign factors 
$\chi^{S}_{I}$. 
\begin{figure} 
\begin{center}
\begin{tabular*}{\textwidth}{c@{\extracolsep{\fill}}
c@{\extracolsep{\fill}}}
\pic{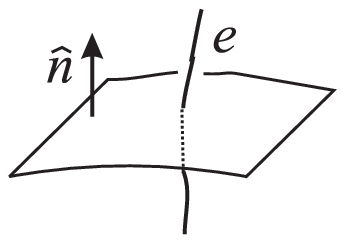}{50}{-15}{-1}{2}
 & 
\pic{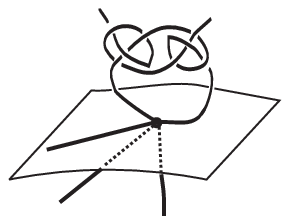}{50}{-15}{-1}{2}\\
(a.) &  (b.) 
\end{tabular*}
\end{center}
\label{figarea} \caption{ Two types of intersections of a spin network 
with a surface (a.) One isolated edge $e$ intersects the surface 
transversely.  The normal $\hat{n}$ is also shown.
(b.)  One vertex of a spin network lies in the surface.  Only 
the non-tangent edges contribute to the area.}
\end{figure}
One can calculate the action of the operator $\hat{O}$ on
an edge $e$ labeled by $n$ as depicted in 
Figure \ref{figarea}(a.).  In this case
the hands act on the same edge so the sign is $1$, 
$\left(\chi^{S}_{I}\right)^{2} =1$,
and the operator becomes $\hat{J}^2$.  The calculation 
makes use of the Pauli matrix identity of Eq. (\ref{pauli}) 
\begin{equation*}
	\begin{split}
		\hat{O}_{e} \ket{s}
		&\equiv  (4 \pi G)^{2} \hat{J}^2 \ket{s} \\
		&= - l^4 \, \frac{n^{2}}{2} \, \kd{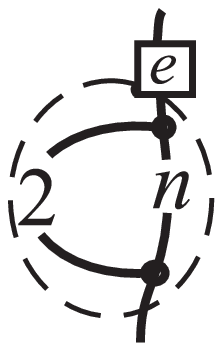} \ket{(s-e)} \\
		& = l^4 \, \frac{n(n+2)}{4} \ket{s}.	
	\end{split}
\end{equation*}
On the second line the edge is shown in the the diagram so it is removed  
from the spin network $s$ giving the state $\ket{(s-e)}$. The diagram is 
reduced using recoupling identities as in Ref. \cite{KL}.
The area operator is the square root of this operator acting at all 
intersections $i$
\begin{equation}
	\hat{A}(S) \ket{s} := \sum_{i} 
	\hat{O_{i}}^{1/2} \ket{s}.
\end{equation}
Thus, the area coming from all the transverse 
edges is \cite{RSareavol}
\begin{equation}	
	\hat{A}(S) \ket{s} = l^{2} \sum_{i}
	\sqrt{ \frac{n_{i}(n_{i}+2)}{4} } \ket{s}.
\end{equation}
The units are collected into the length 
$l \sim 10^{-35}$  m.   The result is 
also easily re-expressed in terms of the 
more familiar angular momentum variables $j=\tfrac{n}{2}$.

The full spectrum of the area operator is found by considering all the 
intersections of the spin network with the surface $S$, including 
vertices which lie on the surface as in Figure \ref{figarea}(b.).  The 
edges incident to a vertex on the surface are divided into three 
categories, those which are above the surface $j^{u}$, below the 
surface $j^{d}$, and tangent to the surface $j^{t}$.  Summing over all 
contributions \cite{QGI}, \cite{FLR}
\begin{equation}
	\label{areaspec}
\hat{A}(S) \ket{s} = \frac{l^{2}}{2} \sum_{v} \left[
2 j_{v}^{u}(j_{v}^{u} +1) + 2 j_{v}^{d}(j_{v}^{d} +1) -
j_{v}^{t}(j_{v}^{t} +1) \right]^{1/2} \ket{s}.
\end{equation}
This result suggests that space is discrete; measurements of area can only
take quantized values.  This property is also seen in the operators 
for quantized directions.

\section{A Cosine Operator}
\label{aop}

The definition of the (first) cosine operator will proceed as in 
Section \ref{spingeom} by first introducing a combinatorial scalar 
product operator and then defining the normalized scalar product or 
cosine operator.  It turns out that, though the combinatorics of both 
operators is perfectly well defined, the classical limit of the scalar 
product operator is singular.  This is expected as, in this approach, 
the metric is ill-defined.  The operator is analogous to the original 
operator introduced by Penrose in that the action of the operator is 
found by attaching a 2-line to two incident edges of a vertex.  The 
precise meaning of the operator, however, only becomes clear when 
``incident edges of the vertex'' are specified.  Further, though the 
operators are similar, the interpretation is not.  The quantum gravity 
operator is a scalar product density.

\subsection{Scalar Product Operator}

The scalar product operator is motivated from the definition of 
$\hat{T}^{(kl)}$.  For simplicity, consider the ``point-wise'' version 
of the operator which measures the scalar product at one vertex of a 
spin network basis state.  As reviewed in Section \ref{qg}, the triad 
operators are expressed in terms of surfaces.  Thus, the scalar 
product operator is associated to two surfaces, instead of two edges 
of a spin network.  The scalar product density operator is defined as
\begin{equation}
	\label{Top}
	\begin{split}
		\hat{T}^{(S_{k}S_{l})}_{v} 
		&:= - \xi \, \chi^{S_{k}}_{I} \chi^{S_{l}}_{I'}
		\kd{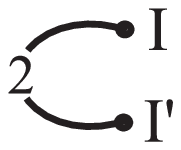} \\
		& \equiv - \xi \chi^{S_{k}}_{I} 
	X_{I}^{j} \; \chi^{S_{l}}_{I'} X^{j}_{I'} 
	\end{split}
\end{equation}
in which $\xi$ is a dimensionful parameter to be fixed by comparison 
to the classical theory.  It will be convenient to divide the edges 
into categories according to their relation to the two surfaces.  They 
are grouped in five categories, labeled by four ``quadrants'' I, II, 
III, and IV, defined by the surface normals and one ``tangent'' $t$ 
for the tangent edges as indicated in Figure \ref{quad}(a.).  The 
interpretation of the quadrants is slightly different than one might 
expect.  An edge is ``in'' a quadrant not when it passes through the 
quadrant, but when the orientation is pointed in the quadrant; if all 
incident edges are outgoing then the categories determine in which 
quadrant the edges lie.

\subsection{Spectrum of the scalar product operator}
\label{SPspectrum}

\begin{figure} 
\begin{center}
\begin{tabular*}{\textwidth}{c@{\extracolsep{\fill}}
c@{\extracolsep{\fill}}}
\pic{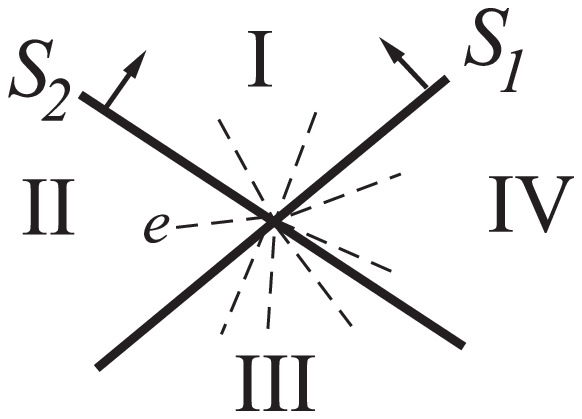}{50}{-15}{-1}{2}
 & 
\pic{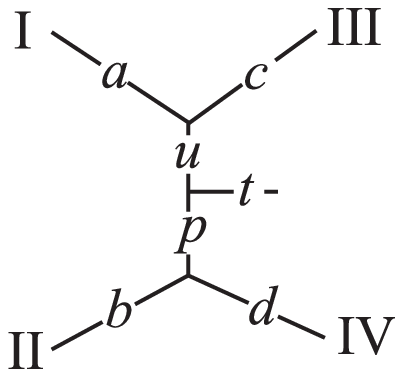}{50}{-15}{-1}{2}\\
(a.) &  (b.) 
\end{tabular*}
\end{center}
		\caption{ \label{quad} (a.)  The intersection of the two 
		surfaces $S_{1}$ and $S_{2}$ with one dimension suppressed.  
		The edges of the graph, when oriented pointing away from the 
		vertex, may be divided into four categories according to where 
		they lie in relation to the two surfaces.  (b.)  The ``core'' 
		of the intertwiner for the vertex.The labels on the lines 
		indicate the representations in each of the four quadrants.  
		The rest of the intertwiner is left unspecified.}
\end{figure}

I present two calculations of the spectrum.  In the spin network basis 
I use the diagrammatic method to find the spectrum.  Then the angular 
momentum operator expression for the scalar product is given.  The 
results are identical.

The operator defined in Eq.  (\ref{Top}) acts on every edge at the 
vertex $v$.  In terms of the diagrammatics, the operator returns the 
state with a 2-line attached.  The original state is recovered after 
simplifying the state using recoupling.  The eigenvalue is determined 
by the sign factor, edge labels and recoupling.

It is useful to choose the intertwiners on the vertex as shown in Fig.  
\ref{quad}(b.).  The edges in the separate quadrants are combined into 
separate internal edges, e.g., the edge labeled with $a$ is the 
combined total of all the edges in ``quadrant I.'' These internal 
edges are then combined, I with III to make $u$ and II with IV to make 
$p$.  Finally, the tangents are included in the internal edge $t$.  
This intertwiner is useful on account of two facts.  First, the 
rotational invariance of the trivalent vertex means that
\begin{equation}
\label{graspslide}
a \, \kd{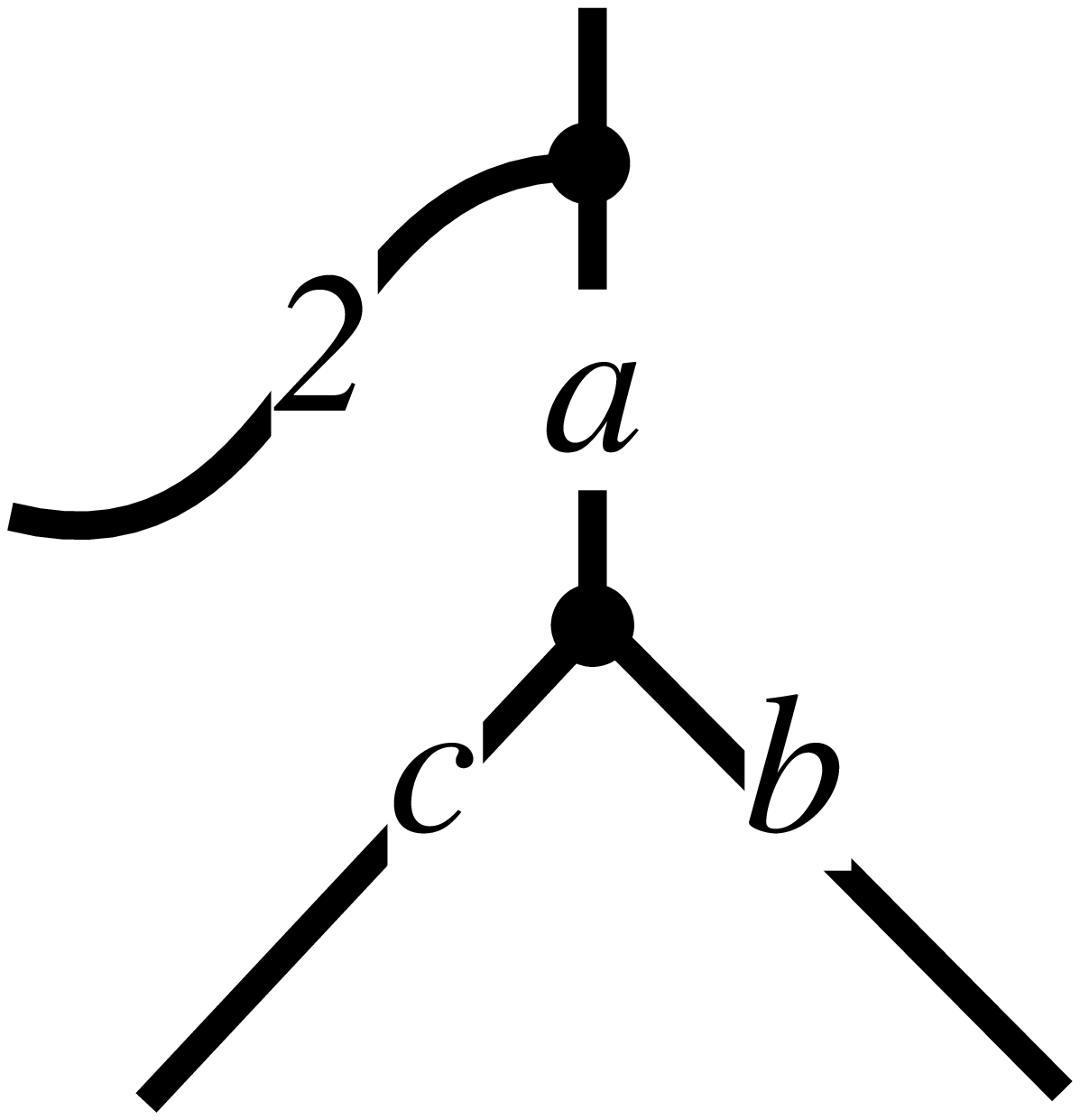} = b \ \kd{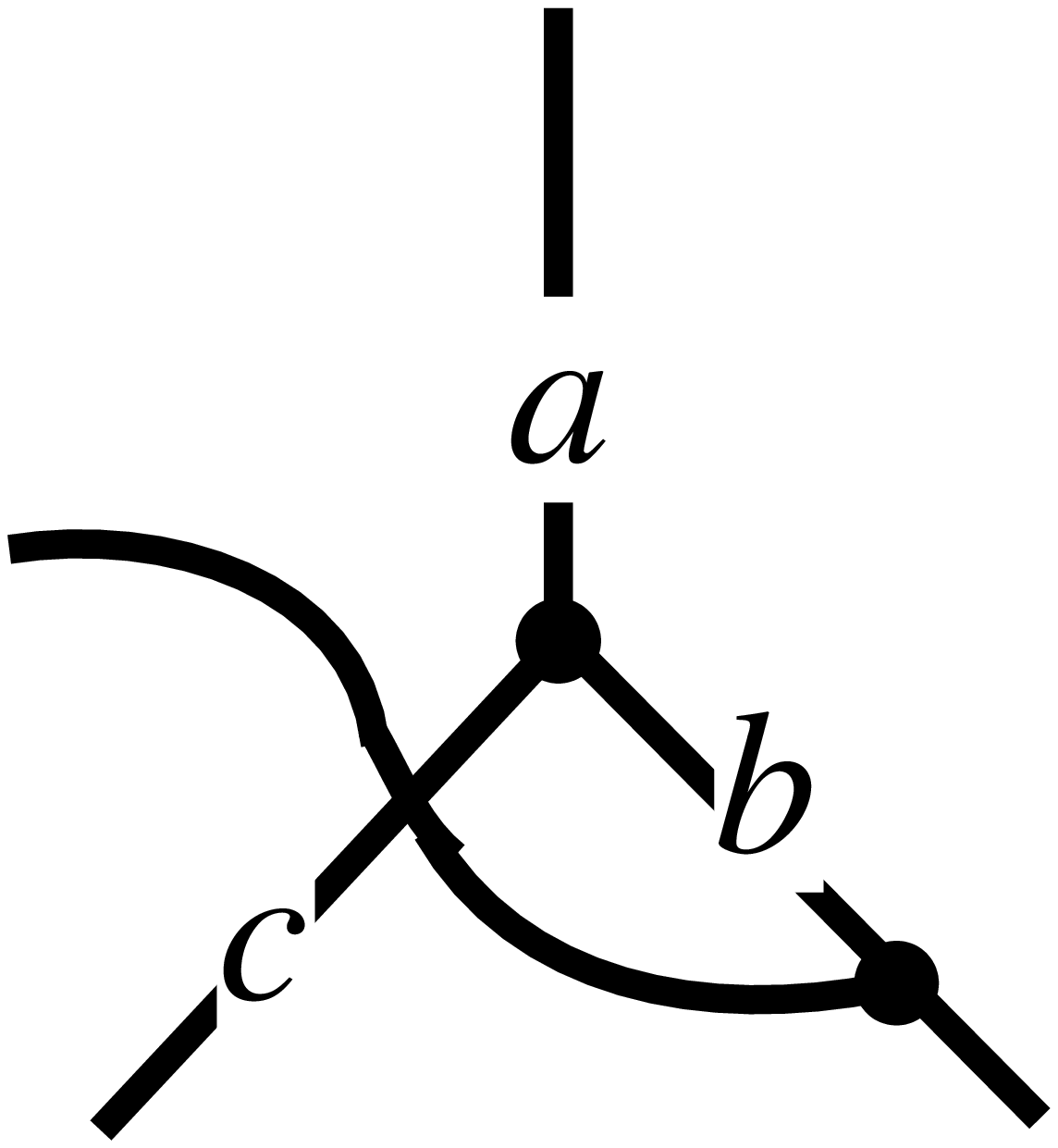} + c \ \kd{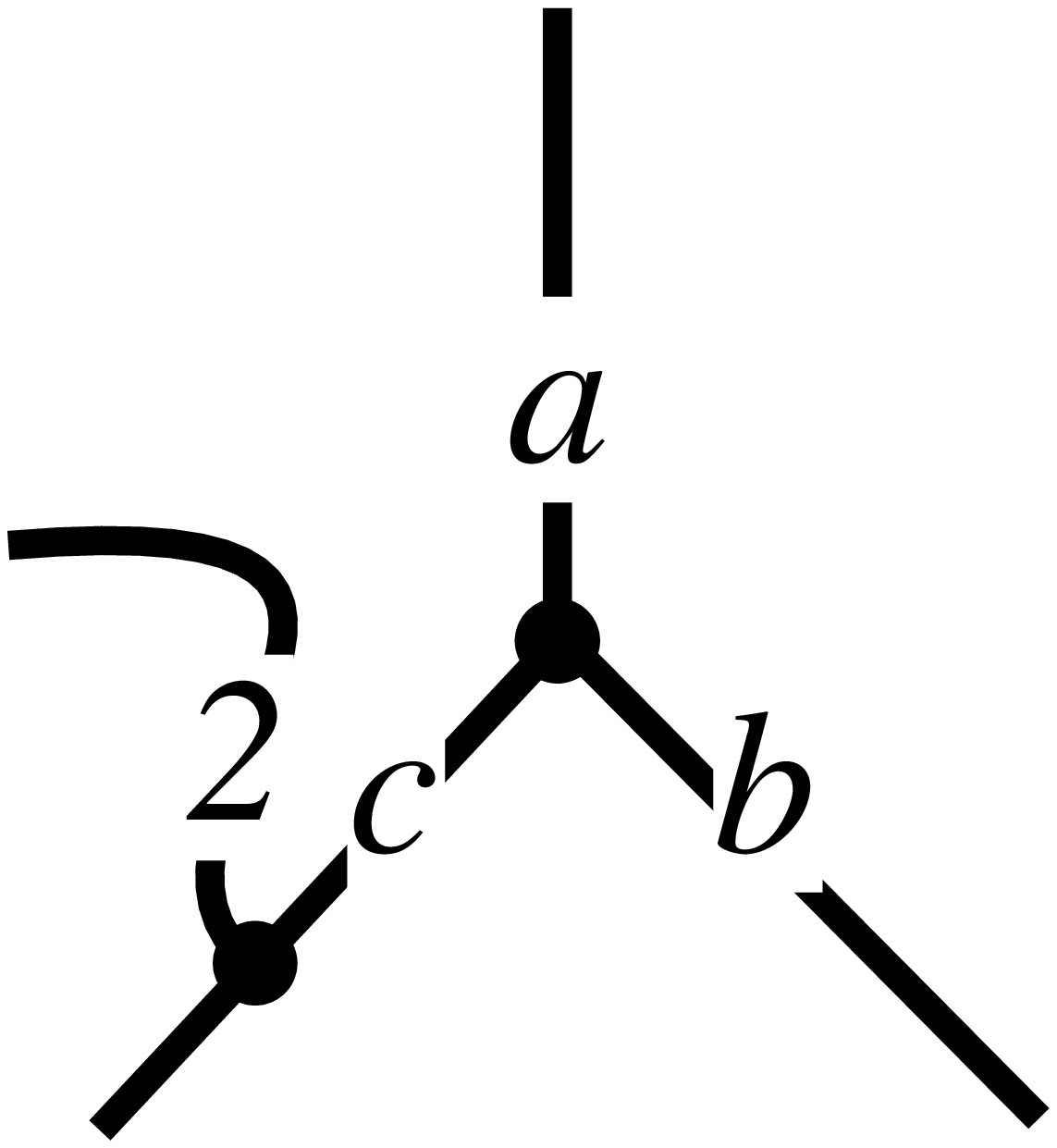}.
\end{equation}
One can ``slide'' the graspings of the incident edges ``down'' to the 
principle internal edges.  (This identity is derived using recoupling 
theory in Ref.  \cite{FLR}.)  Second, the ``cross terms'' cancel, e.g. 
in the notation of Figure (\ref{quad}) for every term with the $S_{1}$ 
hand grasping an edge in the IIIrd quadrant and the $S_{2}$ hand 
grasping an edge in the IInd quadrant, there is an identical term with 
the opposite sign in which the hands grasp the other edges.

The operator acting on a spin network state $s$ then gives, 
with recoupling coefficients $\lambda$, 
\begin{equation}
\label{spec}
	\hat{T}^{(S_{k}S_{l})}_v \ket{s} 
	= \frac{\xi}{2} \left( a^{2} \lambda_{a} + c^{2} 
	\lambda_{c} + 2ac \lambda_{ac} - b^{2} \lambda_{b} - d^{2} 
	\lambda_{d} - 2bd \, \lambda_{bd} \right) \ket{s}.
\end{equation}
The recoupling coefficients come in two types:
\[
\lkd{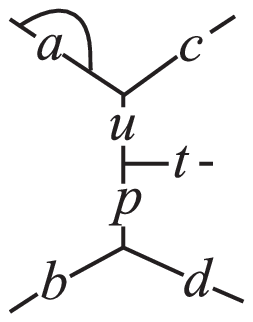} = \lambda_{a} \lkd{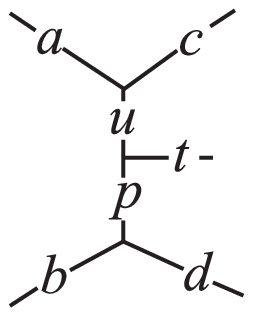}
\]
and
\[
\lkd{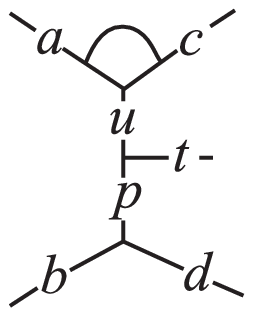} = \lambda_{ac} \lkd{int2}.
\]
These are evaluated to be
\begin{equation}
\lkd{agrasp} = \frac{\theta(a,a,2)}{\Delta_{a}} \lkd{int2}
= - \frac{a+2}{2a}
\lkd{int2}
\end{equation}
and
\begin{equation}
	\begin{split}
	\lkd{acgrasp} & = \frac{ {\rm Tet} 
	\left[ \begin{array}{ccc} a & a & u \\ c & c & 2 \end{array} \right]}
	{\theta(a,c,u)} \lkd{int2} \\
	&= - \frac{a(a+2) + c(c+2) - u(u+2)}{4ac} \lkd{int2}.
\end{split}
\end{equation}
The recoupling quantities may be found in Refs. \cite{FLR}, 
\cite{RC}, and \cite{KL}.  
Substituting these results into Eq. (\ref{spec}) one finds the spectrum of 
the scalar product operator
\begin{equation}
	\begin{split}
	\label{spectrum}
	\hat{T}^{(S_{k}S_{l})} \ket{s} 
	= \frac{\xi}{4} & \left[ 2a(a+2) + 2c(c+2) - u(u+2)  \right.  \\
	& \left. - \left( 2b(b+2) + 2d(d+2) - p(p+2) \right) \right] \ket{s}.
	\end{split}
\end{equation}

The form of this operator immediately implies a number of results.  
Before giving those, however, it is worth deriving this spectrum with 
angular momentum operators.

One may associate one of these operators to each of the quadrants.  
For an $n$-valent vertex the edges are partitioned into those which 
``point'' into the four quadrants and those which are tangent.  Let 
the first $a$ edges be associated to quadrant I, edges $a+1$ to $q$ be 
associated to the IInd quadrant, and so on.  Using the definitions
\begin{equation}
	\begin{split}
\hat{J}^{i}_{(I)} &:= i \hbar \left( X^{i}_{1} + \dots + X^{i}_{a} \right) \\
\hat{J}^{i}_{(II)} &:= i \hbar \left( X^{i}_{a+1} + \dots + X^{i}_{q} \right)\\
\hat{J}^{i}_{(III)} &:= i \hbar \left( X^{i}_{q+1} + \dots + X^{i}_{r} \right)\\
\hat{J}^{i}_{(IV)} &:= i \hbar \left( X^{i}_{r+1} + \dots + X^{i}_{s} \right)\\
\hat{J}^{i}_{(t)} &:=  i \hbar \left( X^{i}_{s+1} + \dots + X^{i}_{n}\right)
	\end{split}
\end{equation}
and the usual rules for angular momentum operators, one may show that
\begin{equation}
	\hat{T}^{(S_{k}S_{l})} 
	= \frac{\xi}{\hbar^{2}} \left[ 2 \hat{J}^{2}_{(I)} + 
	2 \hat{J}^{2}_{(III)} - \hat{J}^{2}_{(I+III)} 
	-2 \hat{J}^{2}_{(II)} - 2 \hat{J}^{2}_{(IV)} + \hat{J}^{2}_{(II+IV)} \right].
\end{equation}
Here, $\hat{J}_{(I+III)} = \hat{J}_{(I)} + \hat{J}_{(III)}$.  
From this expression the 
spectrum may be computed to be
\[
\xi \left[
2 \tfrac{a}{2} (\tfrac{a}{2} + 1) + 2 \tfrac{c}{2} (\tfrac{c}{2} + 
1) -  \tfrac{u}{2} (\tfrac{u}{2} + 1) - 2 \tfrac{b}{2} (\tfrac{b}{2} + 1)
- 2 \tfrac{d}{2} (\tfrac{d}{2} + 1) +  \tfrac{p}{2} (\tfrac{p}{2} + 1)
\right]
\]
as before in Eq. (\ref{spectrum}). 

Now the remarks: (i.)  The operator vanishes when the surfaces do not 
intersect and when there are no vertices in the intersection.  (ii.)  
While the presentation only concerns two surfaces, it is clear that 
this operator is well-defined for all pairs of surfaces with the 
vertex in the intersection.  The spectrum would only involve  
a change in the edge labels $a,b,c,$ and $d$.  This operator, like the spin 
geometry operator, gives an $N \times N$ matrix of scalar products 
(for $N$ surfaces).  (iii.)  By the argument in Ref.  \cite{QGI} the 
operator is densely defined and essentially self-adjoint on ${\cal 
H}_{\mathsf G}$.  (iv.)  This is the complete spectrum.  Briefly, 
suppose to the contrary that a continuous part of the spectrum exists 
then we can project onto this space.  But since $\hat{T}^{(S_k S_l)}$ 
sends ${\rm Cyl}^{2}_{\mathsf G}$ into itself and Cyl is dense in the 
Hilbert space, the projection vanishes (This is an argument given in 
Ref.  \cite{QGI} for the area operator.  See also \cite{ALProj}.)  
(v.)  Finally, there is another form of the scalar product operator 
which makes it formally resemble the scalar product operator in the 
spin geometry theorem.  Defining
\begin{equation}
	\begin{split}
\hat{J}^{i}_{(A_{1})} &:= \hat{J}^{i}_{(I)} + \hat{J}^{i}_{(II)} - 
\hat{J}^{i}_{(III)} - \hat{J}^{i}_{(IV)}, \\
\hat{J}^{i}_{(A_{2})} &:= \hat{J}^{i}_{(I)} - \hat{J}^{i}_{(II)} - 
\hat{J}^{i}_{(III)} + \hat{J}^{i}_{(IV)}, {\rm   and }\\
\hat{J}^{i}_{(T)} &:= \hat{J}^{i}_{(A_{1})} + \hat{J}^{i}_{(A_{2})},
	\end{split}
\end{equation}
one has that the scalar product operator is 
\begin{equation}
	\hat{T}^{(12)} =  \frac{1}{2} \left( \hat{J}^{2}_{(T)} - 
	\hat{J}^{2}_{(A_{1})} - \hat{J}^{2}_{(A_{2})} \right).
\end{equation}
This is formally identical to the operator of Eq.  (\ref{T12op}) used 
in the spin geometry theorem.

\subsection{The cosine operator: Definition}

A cosine can be constructed from the scalar product by normalization.  
The scalar product operator is normalized by the contribution of the 
single vertex $v$ to the areas of both surfaces, Eq.  
(\ref{areaspec}).  This results in a point-wise operator.

Denoting the vertex area 
operators as $\hat{A}_{v}^{S}$, the operator $\hat{C}^{(S_{k}S_{l})}_{v}$ 
is defined as
\begin{equation}
	\label{Copa}
	\begin {split}
	\hat{C}_{v}(S_{k}S_{l}) &:= \frac{1}{\hat{A}_{v}^{S_{k}}}
	\hat{T}^{(S_{k}S_{l})}_{v} 
	\frac{1}{\hat{A}_{v}^{S_{l}}} \\
	& \equiv \frac{1}{\sqrt{ \sum_{I_{v}} \chi^{S_{k}}_{I} 
	X_{I}^{j} \chi^{S_{k}}_{I} X^{j}_{I}}} 
	\sum_{I_{v}, J_{v}} \chi^{S_{k}}_{I} X_{I}^{j} \; 
	\chi^{S_{l}}_{J} X^{j}_{J}
	\frac{1}{\sqrt{ \sum_{I_{v}} \chi^{S_{l}}_{I} 
	X_{I}^{j} \chi^{S_{l}}_{I} X^{j}_{I}}}.
	\end{split}
\end{equation}
As is clear from the non-commutivity of the area operators themselves 
\cite{QGIII}, this $\hat{C}_{v}(S_{k}S_{l})$ operator has ordering 
ambiguities.  On a given vertex, the three operators from which 
$\hat{C}_{v}(S_{k}S_{l})$ is constructed may not commute.  To define a 
cosine operator, it is best to satisfy two properties: There ought to 
be a definite ordering prescription and the operator ought to be 
symmetric.  These two criteria suggest the definition of the first 
cosine operator as
\begin{equation}
	\label{Cop}
	{\rm \hat{C}os}_{v}(S_{k}S_{l}) := \frac{1}{2} \left[
	\hat{C}(S_{k}S_{l})_{v} + 
	\left( \hat{C}(S_{k}S_{l})_{v} \right)^{\dagger} \right].
	\end{equation}
Since the scalar product and area operators are essentially 
self-adjoint, this definition is a simple average of two orderings of 
the operator, $\hat{C}_{v}(S_{k}S_{l})$ and $\hat{C}_{v}(S_{l}S_{k})$.  
The operator of Eq.  (\ref{Cop}) describes the cosine of the angle 
between the two surfaces $S_{k}$ and $S_{l}$ -- the angle between the 
two normals.  In the cases in which the surfaces coincide, it has a 
minimum value $-1$ for identical but oppositely oriented surfaces and 
a maximum value $+1$ when $S_{k} = S_{l}$.  In addition, as the next 
subsection shows, this cosine operator has the correct naive classical 
limit.  However, this operator is cumbersome and may be regarded as 
one explicit construction rather than a definitive definition.

\subsection{The naive classical expression}
\label{classlim}

In this subsection a calculation shows that the cosine operator of Eq.  
(\ref{Cop}), when written in terms of the new variables, has the 
expected form.  Since the ordering issue is a quantum ambiguity, in 
the classical expressions it is not necessary to distinguish between 
${\rm \hat{C}os}_{v}(S_{k}S_{l})$ and $\hat{C}_{v}(S_{k}S_{l})$.

Expressed as a function of the triads the cosine operator becomes
\[
{\rm Cos}_v(S_{k}S_{l})= \frac{E_{S_{k}}^{i} E_{S_{l}}^{i}}
{\sqrt{E_{S_{k}}^{i} E_{S_{k}}^{i}} \sqrt{E_{S_{l}}^{j} E_{S_{l}}^{j}}}.
\]
As the operator only acts in a small region around the vertex, the 
integrals in the definition of $E_{S}^{i}$, Eq.  (\ref{Edef}) may be 
well approximated as
\[
\int d^{2}\tau \, n_{a} \, E^{ai} \approx \Delta^{2}\tau n_{a} E^{ai}.
\]
In this small region one has, when the two unit normals for $S_{k}$ and  
$S_{l}$ are $n$ and $m$, respectively,
\begin{equation}
	\label{classlimcos}
	\begin{split}
	{\rm Cos}_{v}(S_{k}S_{l}) &= 
	\frac{ n_{a} E^{ai} m_{b} E^{bi} }
	{\sqrt{n_{a} E^{ai} n_{b} E^{bi} }\sqrt{m_{a} E^{aj} m_{b} E^{bj}}}
	= \frac{ q \, q^{ab} n_{a} m_{b} }
	{\sqrt{ q q^{ab} n_{a} n_{b} } \, 
	\sqrt{q q^{cd} m_{c} m_{d}}}
	\\ &\equiv cos(\theta)
	\end{split}
\end{equation}
where $\theta$ is the angle between $n_{a}$ and $m_{b}$ and $q^{ab}$ is the 
inverse spatial metric.

\subsection{Operator regularization: Loop and connection representation}
\label{reg}

The scalar product operator is very similar to the area operator of 
quantum geometry.  The same regularization techniques used for the 
area operator can be carried over to the scalar operator case with 
only minor changes.  Therefore, I only sketch the two regularizations.

In the loop representation the regularization of the area observable 
satisfies two properties.
First, when the classical observable is ``pre-regularized,'' 
the classical regularized quantity 
is required to converge to the classical observable.  Second, the 
regularization is required to preserve the invariances of the theory.

In the ``box regularization'' of the area operator \cite{FLR}, the classical 
area observable is first re-expressed as a regularized quantity.  
The surface is partitioned into squares and   
thickened to a three dimensional region.  The two triads are expressed 
in terms of the two-handed loop variable.  The 
classical regularized expression is then the integral of the variable 
over the boxes.  A similar procedure works for the scalar product 
density.

The classical expression to regularize is the scalar product 
density associated to two surfaces
\begin{equation}
	\label{TE}
T^{(S_{k}S_{l})} =  \int_{S_{l}} \int_{S_{k}} d^{2}\sigma d^{2} \tau
\, n_{a}(x(\sigma)) m_{b}(x(\tau)) E^{ai}(x(\sigma)) E^{bi}(x(\tau))
\end{equation}
where $n$ and $m$ are the surface normals for $S_{k}$ and $S_{l}$, 
respectively.  To provide a classical regularized expression for the 
scalar product, one partitions the two surfaces into squares with 
sides $\epsilon_{i}$ for $S_{i}$.  Each surface is then thickened to a 
box $B$ of height $\delta_{i}$.  The dimensions of the boxes are 
linked together $\delta_{l} = \epsilon_{l}^{r}$, $1<r<2$ to give a one 
parameter limit as in Ref.  \cite{FLR}.  The key difference here is 
that two limits must be taken, one for each surface.  The scalar 
product density is only defined in a region around the intersection of 
the two surfaces.  In order that the limits be well-defined the 
partitions of the surfaces are adapted so that the intersection of the 
surfaces lies in the interior of a set of boxes.

Like the area observable the scalar product density may also be 
regularized by the two handed loop variable \cite{RSloop}
\[
\hat{T}^{ab}_{\alpha}(x,y) := - {\rm Tr} \left[ E^{a}(x) U_{\alpha} 
(x,y) E^{b}(y) U_{\alpha}(y,x) \right]
\]
which acts at two points $x$ and $y$.  The classical regularized 
expression is integrated over the thickened surfaces
\[
T^{(S_{k}S_{l})}_{\epsilon} = \frac{1}{2 \delta_{k}\delta_{l}}
\sum_{B_{l}, B_{k}} \int_{B_{k}\otimes B_{l}} 
d^{3}x \, d^{3}y  \; n_{a}(x) \, m_{b}(y) T^{ab}_{\alpha} (x,y).
\]
To lowest order, this is Eq. (\ref{TE}). 

The quantum operator is just the expression with the operator form 
of $T^{ab}_\alpha$.  When the quantum operator acts on a spin network 
edge $e$, the result is
\begin{equation}
\hat{T}^{(S_{k}S_{l})}_{\epsilon} \ket{\kd{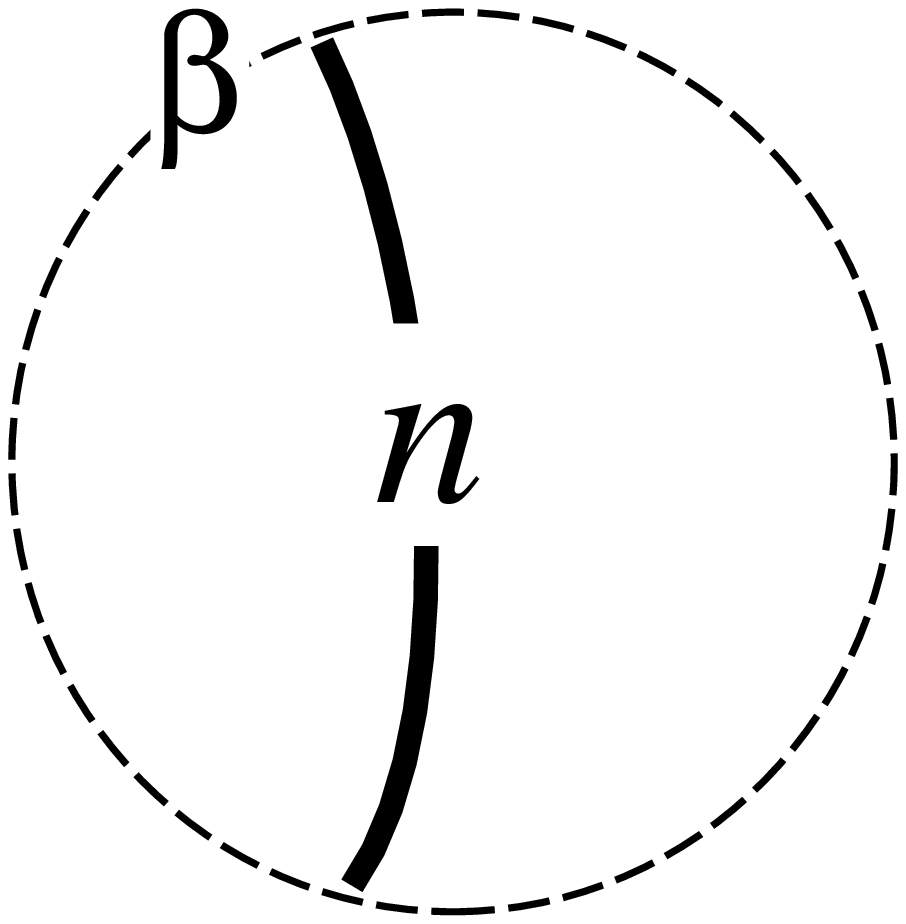}} = l^4 { n^2   \over 2 
\delta^2 } \int_{{B}_k \otimes {B}_l} d^3x \, d^3y \, n_a(x) m_b(y) 
\Delta^a[e,x] \Delta^b[e,y] \, \ket{\kd{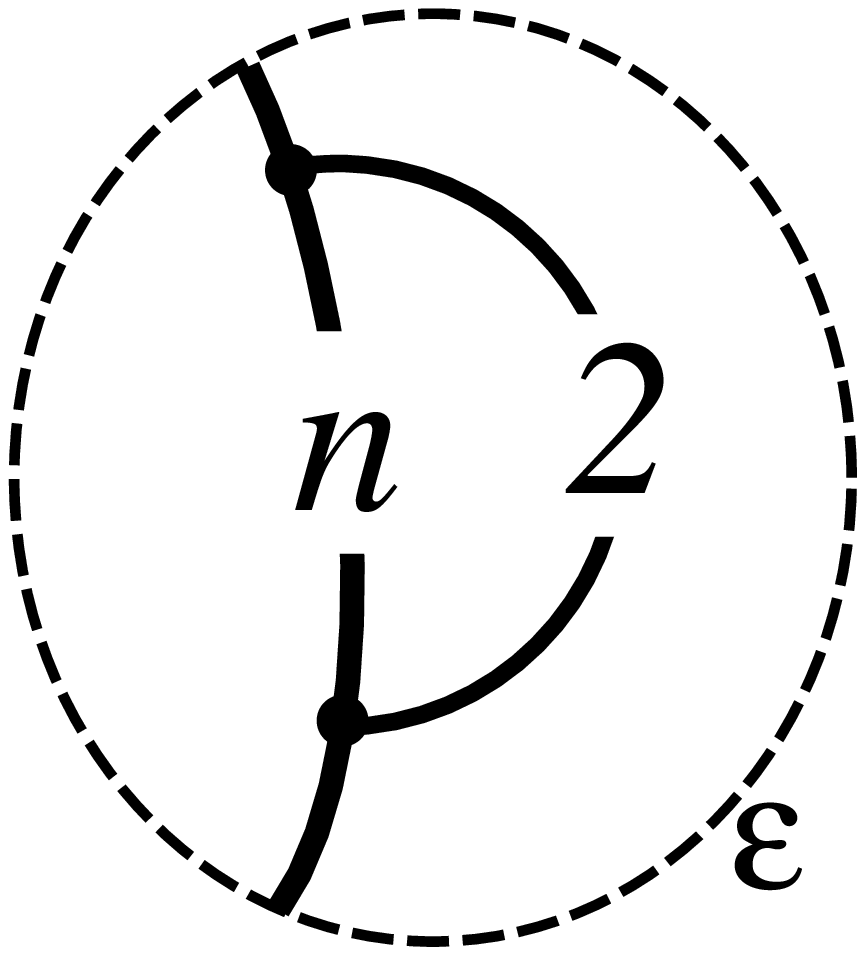}}. 
\end{equation}
The $\hat{T}^{ab}$ operator grasps the edge and the $\epsilon$
indicates that the the limits have yet to be taken.  Letting the
$\delta$-functions eat the spatial integrals, one has
\begin{equation}
l^4 { n^2   \over 2 \delta_{l} \delta_{k} } 
\left( \int_e ds \, n_a(e(s)) \dot{e}^a(s)
\int_e dt \, n_b(e(t)) \dot{e}^b(t) \right) 
\ket{\kd{areae}}.
\end{equation}
In the limit process, which relies on the topology of a continuous 
manifold and not the Hilbert
space,\footnote{The limit cannot be taken in the Hilbert space
topology; it does not exist.  Instead, the limit must be taken in a
topology which remembers this smooth property of the manifold.  The
topology used is induced on the state
space by the classical limit.  That is, a state
$\ket{\alpha_\epsilon}$ converges to the state $\ket{\alpha}$ if
$\alpha_\epsilon$ converges pointwise to $\alpha$.}
the integrals reduce to \cite{FLR} 
\[
\int_e dt \, n_b(e(t)) \dot{e}^b(t)
 = \begin{cases} 0 & \text{if $e$ is tangent to $S$} \\
\pm \delta/2 & \text{otherwise.} \end{cases}
\]
The diagrammatic operator is equivalent to the product of two invariant 
vector fields so we have, in the limit,
\[
\hat{T}^{(S_{k}S_{l})} \ket{s} =  - \frac{l^{4}}{8} 
\chi^{S_{k}}_{I} X^{i}_{I} \; \chi^{S_{l}}_{J} X^{i}_{J}
\]
in which $l^{2} = 4 \pi G \hbar$.  
Comparing this result with the definition Eq. (\ref{Top}), we learn that the 
constant $\xi$ is fixed as
\[
\xi = \frac{l^{4}}{8}.
\]
The classical regularized expression fixes the parameter $\xi$.

The cosine operator may be regularized using the loop regularizations for the 
scalar product operator as above and the area operator as in Ref. 
\cite{FLR}.   

One may also perform a regularization directly with the triads.  The 
chief advantage of this regularization is that the limits exist in the 
Hilbert space topology.  The regularization of the area makes use of 
tempered triad operators integrated over the surface.  Given a 
Lie-algebra valued field on $S$, the classical variables are
\[ 
[E_{S}]_{f} := \int_S d^2 \sigma \, f^i_\epsilon (x(\sigma, v) \, n_a 
E^{ai}(x(\sigma)).
\]
The fields are chosen to be of compact support on $S$.  As $\epsilon$ 
goes to zero, the support of $ f^i_\epsilon (x(\sigma, v)$ contracts 
to the point $v$.  In this version of the regularization one quantizes 
these triads directly, replacing $E^{ai}$ with $-i \hbar \delta / 
\delta A_a^i$.

When the triads are integrated with test functions of compact support. 
This region shrinks to a point as the regulator $\epsilon$ vanishes. 
So the scalar product operator may be expressed as
\begin{equation}
	\begin{split}
[\hat{T}^{(S_{l}S_{k})}]_{f}(x) \cdot \Psi_{\mathsf G} &:=
[E_{S_{l}}]_{f} [E_{S_{k}}]_{f} \cdot \Psi_{\mathsf G} \\
&= - \frac{l^{4}}{4} \sum_{v, v'} \sum_{I,J} \chi^{S_{l}}_{I} 
f_{\epsilon} (x,v)
X^{i}_{I} \, \chi^{S_{k}}_{J} f_{\epsilon}(x, v') X^{i}_{J} \cdot 
\Psi_{\mathsf G}.
\end{split}
\end{equation}
The first sum is over vertices $v,v'$ in the intersection of the two
surfaces and the second is over incident edges.  As $\epsilon$ tends 
to zero, the operators act on a single compact region and the operator 
becomes
\[
[\hat{T}^{(S_{l}S_{k})}_{v}]_{f} = - \frac{l^{4}}{4} \sum_{v} 
\left( f_{\epsilon}(x,v) \right)^{2}  \sum_{I,J } \chi^{S_{l}}_{I} 
X^{i}_{I} \chi^{S_{k}}_{J}  X^{i}_{J}.
\]
This operator is similar to the determinant of the induced surface 
metric operator $[\hat{g}_S]_f$ of Ref.  \cite{QGI}.  Like this metric 
operator, the $[\hat{T}^{(S_{l}S_{k})}_{v}]_{f}$ develops a 
$\delta^4(0)$ singularity in the limit.

However, the cosine operator when regularized in this manner, is well 
defined.  As $\epsilon$ goes to zero, it becomes,
\begin{equation}
	[{\rm \hat{C}os}(S_{l}S_{k})]_{f} = \frac{ 
\left( f_{\epsilon}(x,v) \right)^{2}  \sum_{I,J } \chi^{S_{l}}_{I} 
X^{i}_{I} \chi^{S_{k}}_{I}  X^{i}_{J}}{\sqrt{ \left(
f_{\epsilon}(x,v) \right)^{2} \sum_{I,J } \chi^{S_{l}}_{I} 
X^{i}_{I} \chi^{S_{l}}_{J}  X^{i}_{J}}\, \sqrt{ 
\left( f_{\epsilon}(x,v) \right)^{2}  \sum_{I,J } \chi^{S_{k}}_{I} 
X^{i}_{I} \chi^{S_{k}}_{J}  X^{i}_{J}}}. 
\end{equation}
Since the test functions cancel, this operator is independent of 
$\epsilon$ and so is manifestly well-defined in the limit; $ [{\rm 
\hat{C}os}(S_{l}S_{k})]_{f} \equiv \hat{C}_{v}(S_{l}S_{k})$.  
This result is just the operator of Eq.  (\ref{Copa}).

\section{An angle operator}
\label{jetop}

There is another definition of the angle operator which may be even 
more useful.  In many ways closer to the operator used by Penrose in 
the diagrammatic form of the spin geometry theorem, this operator 
assigns an angle to two bundles of edges incident to a vertex.
Like the scalar product operator, the quantities which determine the 
angle are the ``internal edges'' of a spin network vertex, rather than 
the external edges.
\begin{figure} 
\begin{center}
\begin{tabular*}{\textwidth}{c@{\extracolsep{\fill}}
c@{\extracolsep{\fill}}}
\pic{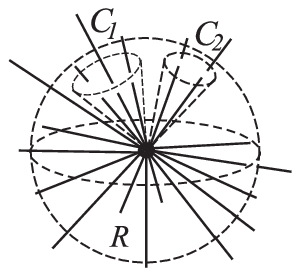}{70}{-10}{-1}{2}
 & 
\pic{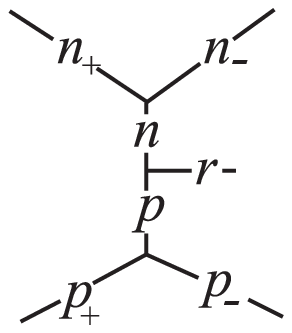}{60}{-10}{-1}{2}\\
(a.) &  (b.) 
\end{tabular*}
\end{center}
		\caption{ \label{cones} (a.)  A vertex with two bundles of edges 
		defined by the cones $C_{1}$ and $C_{2}$. 
		(b.)  The 
		``core'' of the intertwiner for the vertex is chosen so that 
		there are $n$ lines (divided into two internal lines $n_{+}$ 
		and $n_{-}$) in $C_{1}$, $p$ lines in $C_{2}$, and the 
		remainder of the incident edges in $r$.}
\end{figure}

The quantum operator is defined as follows.  Around any single vertex 
incident edges are partitioned into three categories.  I will call 
them $C_{1}$, $C_{2}$, and $R$ (the names are motivated below).  
Associated to these partitions are three spin operators $J_{1}$, 
$J_{2}$, and $J_{r}$ and a trivalent vertex labeled by $n$, $p$ and 
$r$.  The quantum angle operator is defined to be\footnote{I thank 
Roberto DePietri for suggesting the form of this operator 
\cite{robertocom}.}
\begin{equation}
	\label{Aop}
\hat{\theta}^{(12)}_{v} := \arccos \frac{\hat{J}_{1} \cdot \hat{J}_{2}}
{|\hat{J}_{1}| \, | \hat{J}_{2} |}
\end{equation}
in which $|\hat{J}| = \sqrt{\hat{J}^{2}}$.  After the work of Section 
\ref{SPspectrum}, deriving the spectrum is immediate.  The result is
\[
\theta^{(12)}_{v} = {\rm arccos} \left( \frac{j_{r}(j_{r}+1) -
j_{1}(j_{1} +1) - j_{2}(j_{2}+1)} {2 \left[ j_{1}(j_{1} +1 ) \, 
j_{2}(j_{2} +1 ) \right]^{1/2}} \right).
\]

The key idea of the relation to quantum geometry is to measure the 
angle between two bundles of edges each contained within disjoint 
``conical regions'' with the vertices of the cones based at a spin 
network vertex.  As shown in Fig.  \ref{cones}a, I introduce a closed 
surface in the neighborhood of the vertex.  This surface, which is 
topologically $S^{2}$, may be partitioned into three regions, say 
$S_{1}$, $S_{2}$, and $S_{r}$, such that $S_{1}$ and $S_{2}$ are 
simply connected, disjoint regions while $S_{r}$ is multiply connected 
(a sphere with two holes cut out).  For simplicity, the partition is 
made so that no edges intersect the boundaries of the regions.  The 
surfaces $S_{1}$ and $S_{2}$ define the regions for which the angle is 
defined.  By construction, all the intersections between the surfaces 
and the spin network are bivalent.  In the following discussions of 
the naive classical limit and the regularization, it is convenient to 
introduce a flat background metric in the neighborhood of the vertex 
and a ``radius'' of the spherical surface, $\delta$.  As this radius 
varies, the surfaces $S_{1}$ and $S_{2}$ are defined so that they 
always form the base of the cones $C_{1}$ and $C_{2}$.  Thus, the 
edges and representations on the edges intersecting the surfaces do 
not depend on $\delta$.

The naive classical limit is very similar to the cosine operator in 
Section \ref{classlim}.  In fact, the operator has the same form, 
only the definitions of the surfaces has changed.  Using the
above surfaces, the operator
\[
\frac{J_{1} \cdot J_{2}}{|J_{1}| \, 
| J_{2} |}
\]
has the same expression in terms of triads as the cosine operator of
Eq. (\ref{classlimcos}).  Again this operator has the correct behavior 
in the naive classical limit.

The regularization is straight forward, given the surfaces and the 
parameter $\delta$.  The product of triads 
associated to the two surfaces is defined as
\[
\left[ T^{(S_{1}S_{2})}_{v} \right]_{\delta} := 
E^{i}_{S_{1}} \, E^{i}_{S_{2}}.
\]
While in the limit of vanishing $\delta$ this product is not well 
defined, the function
\[
\frac{ \left[ {T}^{(S_{1}S_{2})}_{v}\right]_{\delta}}
{A( S_{1} ) A( S_{2} )},
\]
is well defined in the limit.\footnote{If the nature of this limit 
makes the reader uncomfortable, one can also keep the radius fixed and 
use the regularization parameter to deform the surfaces from circles 
on a sphere to cones with the apex at the spin network vertex.  The 
result is two conical surfaces which intersect the spin network only 
at the vertex.} One may now define the (2nd) cosine operator ${\rm 
\hat{c}os}^{(12)}_{v}$ to be
\[
{\rm \hat{c}os}^{(12)}_{v} := 
\frac{ \left[ \hat{T}^{(S_{1}S_{2})}_{v}\right]_{\delta}}
{A( S_{1} ) A( S_{2} )}
\]
which is equivalent to
\[
\frac{1}{2} 
\frac{\hat{J}^{2}_{r} - \hat{J}^{2}_{1} - \hat{J}_{2}^{2}}
{| \hat{J}_{1} | \, | \hat{J}_{2} |}.
\]  
when the orientations on the edges is taken to be outgoing.

The result of this construction is that the incident edges are neatly 
divided into three categories which are the sums of representations 
passing through the surfaces $S_{1}$, $S_{2}$, and $J_{(r)}$.  So 
there is a natural choice for intertwiner at the vertex $v$ such that 
all the incident edges passing through each surface combine into one 
internal edge.  The ``core'' of the intertwiner is labeled by $j_{1}$, 
$j_{2}$, and $j_{r}$.  The last spin, $j_{r}$, is the remainder of the 
total spin.

At this stage it is worth commenting that all these 
operators {\em commute}.  This is easily seen by noticing that the 
partition of the edges into cones 1 and 2 and ``the rest'' creates 
three disjoint sets of edges.  Also, it is worth noting that, in a 
small region around a vertex, there are no edges tangent to the 
surfaces.  Thus, this operator is well-defined on the Hilbert space.  
The angle operator is defined in the expected way
\[
\hat{\theta}^{(12)}_{v} := {\rm arccos} \left( {\rm \hat{c}os}^{(12)}_{v}
\right).
\]
This angle operator has the nice property that an angle is determined 
quantum-geometrically by the internal structure of the intertwiner at 
the vertex.  This operator is also ``realistic'' in that the angle 
operator compares two regions rather than single edges.

Finally it is worth commenting that the operator is still defined for 
edges of arbitrary orientation, although the spectrum is modified.
In this case, the cosine operator is
\begin{equation}
	\frac{ J^{2}_{(n_{+}+p_{+})} - J^{2}_{(n_{+}+p_{-})} -
	J^{2}_{(n_{-}+p_{+})} +J^{2}_{(n_{-}+p_{-})} }{ 2 \sqrt{J^{2}_{(1)}}
	\sqrt{J^{2}_{(2)}}}
\end{equation}
in which $n_{+}, n_{-}$ ($p_{+}, p_{-}$) label the edges passing 
through the surface $S_{1}$ ($S_{2}$).

\section{Conclusion}

Based on the operators of the spin geometry theorem, several operators 
for quantum geometry were introduced.  The scalar product density 
operator was defined in Eq.  (\ref{Top}).  It only acts on a single 
vertex $v$ of the spin network and is only non-vanishing when $v$ lies 
in the intersection of the two surfaces $S_{l}$ and $S_{k}$.  The 
overall parameter was fixed by the classical limit in Section 
\ref{reg}, giving the result
\begin{equation}
	\hat{T}^{(S_{k}S_{l})}_{v} 
		:= - \frac{l^{4}}{8}  \, \chi^{S_{k}}_{I} \chi^{S_{l}}_{I'}
		\kd{spop}.
\end{equation}
The bounded and discrete spectrum is given in Eq.  (\ref{spectrum}).  
Though classically ill-behaved, this is a well-defined quantum 
operator.

The more complex cosine operator was 
defined in Eq. (\ref{Cop}) with the ordering
\[
	{\rm \hat{C}os}_{v}(S_{k}S_{l}) := \frac{1}{2} \left[
	\hat{C}^{(S_{k}S_{l})}_{v} + 
	\left( \hat{C}^{(S_{k}S_{l})}_{v} \right)^{\dagger} \right].
\]
Closely related to the operator of the spin geometry theorem, this 
operator measures the cosine of the angle between the normals of two 
surfaces.  It is only non-vanishing at the intersection of these 
surfaces and at the vertices of a spin network state.  While this 
operator has the correct form in the naive classical limit, the 
quantum definition, due to ordering difficulties, is cumbersome.

An operator which measures angles between two ``conical regions'' was 
defined in Section \ref{jetop} via
\[
\hat{\theta}^{(12)}_{v} := \arccos \frac{\hat{J}_{(1)} \cdot \hat{J}_{(2)}}
{\sqrt{\hat{J}_{(1)}^{2}} \, \sqrt{\hat{J}_{(2)}^{2}}}.
\]
This angle operator is simple, quantum mechanically well-behaved, and 
has the correct naive classical limit.

There are several striking properties of these operators.  Both the 
cosine and angle operators have fully discrete spectra and are {\em 
independent} of the Planck scale.  This is easily seen in the 
regularizations; the factors of the length scale $l$ cancel in the 
spectrum of the cosine operator.  While required on dimensional 
grounds, this independence is new to the operators of quantum geometry.  
This property does not indicate that the angle discreteness is 
coarsely grained.  Rather, the grain is determined by the 
(semi)classical state on which the operator acts (see below).  An 
obvious corollary to this is that these operators are the first 
operators of quantum geometry which are free of the Immirizi-parameter 
ambiguity (\cite{I} - \cite{GOPI}) since they contain equal powers of the 
triad operator in the numerator and denominator.  This is an 
operator which is independent of the Immirizi-parameter-sector of the 
Hilbert space.

I close with three remarks on the wider implications of these 
definitions and a discussion of the spin geometry theorem.  (i.)  While 
the operator ${\rm \hat{C}os}_{v}(S_{k}S_{l})$ is gauge invariant and 
diffeomorphism invariant (at least when the surfaces are specified in 
a diffeomorphism covariant manner such as by value of a scalar field), 
there are other closely related operators which may be of interest.  
One possibility for a realistic microscopic operator would be to 
average the value of the cosine operator over a small region defined 
by two thickened surfaces.  This microscopic operator is an average 
over the cosines of all the vertices in the region which would better 
approximate realistic measurements of angles.  A classical observable 
for the cosine averaged over a volume $R$ is constructed from two 
scalar fields $\phi$ and $\pi$,
\[
[\cos \theta](\phi, \pi, R) := \frac{1}{V(R)} \int_{R} d^{3}x  
\frac{ q^{ab} \,  \nabla_{a} \phi \nabla_{b}\pi }
{\left[ q^{ab} \, \nabla_{a} \phi \nabla_{b} \phi \; 
q^{cd} \, \nabla_{c} \pi \nabla_{d} \pi \right]^{1/2}} 
\]
-- effectively a correlation of degrees of freedom.
The corresponding quantum operator might be the volume averaged
\[
\frac{1}{\hat{V}(R)} \sum_{v \in R} {\rm \hat{C}os}_{v}^{S_{l} S_{k}}
\hat{V}_{v}
\]
for level surfaces $S_{l}$ and $S_{k}$ of the scalar fields and 
vertices $v$ in the region $R$.  At this stage, however, it is far 
from clear whether this operator could be well-defined.

Even so, a realistic model of the way in which angles are measured -- 
even on an atomic scale -- would involve averages over large (in terms 
of Planck volumes) spaces, perhaps along the lines indicated above.  
One can envision a network built from simple elements such as 4-valent 
vertices which might be able to support an accurate description of 
angles in a continuous space.

(ii.)  The scalar product operator defined in Eq.  (\ref{Top}) is 
similar to the $\hat{Q}(\omega)$ operator used in the construction of 
the first weave states \cite{ARS}.  There are two key differences.  
First, the $Q$ operator corresponds to the classical quantity
\[
Q[\omega] = \int d^{3}x (\omega_{a} E^{ai}\, \omega_{b} E^{bi})^{1/2}
\]
so $T^{(kk)}$ is ``the square'' of this operator.  Second, the Q 
operator is based on a single one-form while the scalar product operator 
is based on two, distinct surface normals.

(iii.)  One classical property of angles which is not obviously 
preserved on all quantum states by these operators is additivity.  
Classically we expect that the sum of the measurement of two adjacent 
angles equals the measurement of the total angle.  Quantum 
mechanically this is not so.  For instance consider three surfaces: 
two , $S_{1}$ and $S_{2}$, as in Figure (\ref{quad}) and one, $S_{3}$, 
passing through the IInd and IVth quadrants.  If all the edges of the 
spin network in these two quadrants are concentrated in the two wedges 
between the surfaces $S_{1}$ and $S_{3}$ then the angles 
$\theta(S_{1}, S_{2})$ and $\theta(S_{1}, S_{3})$ are identical.  Of 
course, the definition of these surfaces is a more delicate matter.  
It may be that this spin network vertex does not support a geometrical 
interpretation which makes a distinction between these surfaces.

With these angle operators it is possible to use the spin geometry 
theorem to partially characterize semiclassical states.  Though the 
scalar product density operator of Eq.  (\ref{Top}),  
differs from the analogous operator in the spin geometry theorem, some 
results carry over to the quantum geometry context.

For simplicity, I describe the results only for the angle operator of 
Eq.  (\ref{Aop}).  The analysis of the spin geometry theorem uses 
properties of the scalar product which I take to be $\hat{J}_{(1)} 
\cdot \hat{J}_{(2)}$ as in numerator of the definition.  It is a 
well-defined, essentially self-adjoint operator on the Hilbert space 
${\cal H}$.  These properties ensure that the main result of the spin 
geometry theorem holds in the quantum geometry context.

The key piece of the theorem is a relation between the uncertainty in 
the ``normalized'' scalar product operator and the size of 
determinants of four vectors.  The normalized operator is the scalar 
product operator divided by the maximum possible value.  On a fixed 
spin network, since the spectrum of the scalar product operator is
\[
	\tfrac{1}{2} \left[ j_{r}(j_{r}+1) - j_{1}(j_{1}+1) -
	j_{2}(j_{2}+1) \right]
\]
which obtains a maximum when $j_{r}=j_{1}+j_{2}$, the maximum is 
$j_{1}j_{2}$ \cite{JM}.  I denote this normalized scalar product 
operator with $\hat{t}^{(kl)}$.  The uncertainties in the 
operator $\hat{t}^{(kl)}$ are small for large spins.

Further, the uncertainty relations for this operator imply that the 
operator can be $\delta$-classical only if the spins are large.  This 
may be seen using
\[
\ev{ \Delta A} \ev{\Delta B} \geq \frac{1}{2} \left|
\ev{ [ A, B ]} \right|
\]
and
\[
\left[ J_{(1)} \cdot J_{(2)}, J_{(2)} \cdot J_{(3)} \right] = 
i J_{(1)} \cdot J_{(2)} \times J_{(3)}.
\]
A calculation shows that the uncertainty is bounded by
\begin{equation}
	\label{uncertainty}
\delta^{2} > \Delta t_{12} \Delta t_{23} \geq
\frac{A}{2j_{2}}
\end{equation}
where 
\[
A = \frac{|\ev{J_{(1)} \cdot J_{(2)} \times J_{(3)}}|}
{j_{1}j_{2}j_{3}}
\]
is of $O(1)$ and $j_{2}$ is the spin on the internal edge shared by 
both operators \cite{JM}.  Since the spin geometry theorem ties the 
uncertainty in the normalized scalar product operator 
($\delta$-classical) to the approximation of the angle 
($\epsilon$-constraints), angles are precisely defined only when the 
spin on the internal edges is large.  The internal edges are involved 
because these are the only representations that are measured by the 
operators $\hat{J}_{(k)}$.  Alternately one can see this as a result 
of operator grasping all edges passing through a particular surface.  
On account of the identity of Eq.  (\ref{graspslide}), it is only the 
spin of the core intertwiner which is measured.  Thus, the 
$\delta$-classical limit does not require that the external spins be 
large.

Since the scalar product operator is identical to the one used in the 
spin geometry theorem and since the vertex ensures that the spins are 
correlated, in the limit of large spins, the spin network states can 
support the interpretation of angles in three dimensional space.  The 
key step in the proof, which is demonstrated in Ref.  \cite{JM}, is to 
show that for any $\epsilon >0$ there exists $\delta >0$ such that 
when $\Delta \hat{t}^{(kl)} < \delta$ for internal edges $k$ and $l$ 
in a 4-tuple, then the determinant the $4 \times 4$ submatrix of these 
edges is less than $\epsilon$.  When this condition is met than the 
state may be said to support angles via the scalar product 
proposition.  We learn from Eq.  (\ref{uncertainty}) that $\Delta 
\hat{t}^{(kl)}$ is small when the spins on the internal edges is 
large.  Therefore spin network states approximate angles in three 
space with arbitrary accuracy, given sufficiently large internal 
spins.\footnote{In the case of the scalar product operator of Eq.  
(\ref{Top}), the spectrum is not positive, semi-definite.  The scalar 
product lemma of Section \ref{spingeom} would have to be generalized 
for this case.  Alternately, another operator could be defined which 
has a ``shifted spectrum.''}

This argument suggests that the classical limit of spin networks -- a 
state that accurately models the ``continuum'' we see around us -- 
includes a condition on the spins on internal edges.  The larger they 
are, the closer one can come to a continuum of angles.  The condition 
does not specify how the large spins are distributed on external 
lines.  In fact these spins could even be spin 1/2.  Using naive 
statistical arguments it seems plausible that the most likely 
configuration is one with small spins on many external lines -- a 
highly flocculent network.  However, the precise form of the 
conditions on the internal edges is more subtle.  For, the all 
possible internal spins need to be large (corresponding to all pairs 
of surfaces).  But it is not clear how define these surfaces or cones 
``realistically.''  Further, one might suspect that such a vertex, on 
account of the concentration of geometric flux, would be a point of 
high curvature.  This would represent a sharp departure from flat 
space.  Indeed, there is some reason to suspect that this the case 
\cite{SMqqe}.  Clearly, a condition on the internal spins is but one 
condition for the semiclassical limit.  What this condition, together 
with the angle operator, does show is that the spin network states of 
quantum geometry are able to support an interpretation of angles in 
three dimensional space.

\begin{ack} 
It is a pleasure to thank Jos\'e A. Zapata for discussions throughout 
this project and Kirill Krasnov for comments on the manuscript.  I 
also thank Roberto DePietri for a discussion on the matter of a sign 
and, especially, for his invaluable comments which led to the angle 
operator of Section~\ref{jetop}.  I gratefully acknowledge the support 
of the Austrian Science Foundation (FWF) through a Lise Meitner 
Fellowship.
\end{ack}

\end{document}